\documentclass[final,3p,twocolumn]{elsarticle}

\usepackage{amsmath}
\usepackage{txfonts}
\usepackage[utf8x]{inputenc}
\usepackage{graphicx}
\usepackage[protrusion=true,expansion=true]{microtype}
\usepackage{hyperref,xcolor}
\usepackage[T1]{fontenc} 
\usepackage{graphicx}
\usepackage{ulem}

\journal{Astroparticle Physics}

\begin{document}

\begin{frontmatter}

\title{Radio Morphing: \\towards a fast computation of the radio signal from air showers}

\author[a]{Anne Zilles}
\ead{zilles@iap.fr}
\author[b,c]{Olivier Martineau-Huynh}
\author[a]{Kumiko Kotera}
\author[g]{Matias Tueros}
\author[e]{Krijn de Vries}
\author[d]{Washington Carvalho Jr.}
\author[f]{Valentin Niess}
\author[a]{Nicolas Renault-Tinacci}
\author[a]{Valentin Decoene}

\address[a]{Sorbonne Universit\'{e}, UPMC Univ.  Paris 6 et CNRS, UMR 7095, Institut d'Astrophysique de Paris, 98 bis bd Arago, 75014 Paris, France}
\address[b]{LPNHE, CNRS-IN2P3 et Universit\'{e}s Paris VI \& VII, 4 place Jussieu, 75252 Paris, France}
\address[c]{National Astronomical Observatories, Chinese Academy of Sciences, Beijing 100012, China}
\address[d]{Universidade de Santiago de Compostela, 15782 Santiago de Compostela, Spain}
\address[e]{Vrije Universiteit Brussel, Physics Department, Pleinlaan 2, 1050 Brussels, Belgium}
\address[f]{Clermont Universit\'{e}, Universit\'{e} Blaise Pascal, CNRS/IN2P3, Laboratoire de Physique Corpusculaire, BP. 10448, 63000 Clermond-Ferrand, France}
\address[g]{Instituto de F\'{i}sica La Plata - CONICET/CCT- La Plata. Calle 49 esq 115. La Plata, Buenos Aires, Argentina}

\date{Received $\langle$date$\rangle$ / Accepted $\langle$date$\rangle$}

\begin{abstract}
Over the last years, radio detection has matured to become a competitive method for the detection of air showers. Arrays of thousands of antennas are now envisioned for the detection of cosmic rays of ultra high energy or neutrinos of astrophysical origin. The data exploitation of such detectors requires to run massive air-shower simulations to evaluate the radio signal at each antenna position. In order to reduce the associated computational cost, we have developed a semi-analytical method for the computation of the emitted radio signal called {\it Radio Morphing}. The method consists in computing the radio signal of any air-shower at any location from the simulation of one single reference shower at given positions by i) a scaling of the electric-field amplitude of this reference shower, ii) an isometry on the simulated positions and iii) an interpolation of the radio pulse at the desired position. This technique enables one to compute electric field time traces with characteristics very similar to those obtained with standard computation methods, but with computation times reduced by several orders of magnitude. In this paper, we present this novel tool, explain its methodology, and discuss its limitations. 
Furthermore, we validate the method on a typical event set for the future GRAND experiment showing that the calculated peak amplitudes are consistent with the results from ZHAireS simulations with a mean offset of $+8.5$\% and a standard deviation of $27.2\%$ in this specific case. This overestimation of the signal strength by Radio Morphing arises mainly from the choice of the underlying reference shower.
\end{abstract}

\begin{keyword}
radio detection, high-energy astroparticles, air-shower simulation, radio-signal parametrization\\
\vspace{0.5cm}
\noindent
{\small \textcopyright  2019. This manuscript version is made available under the CC-BY-NC-ND 4.0 license \url{http://creativecommons.org/licenses/by-nc-nd/4.0/}}
\end{keyword}

\end{frontmatter}



\section{Introduction}
An extensive air-shower develops when a primary high-energy cosmic particle interacts with molecules in the atmosphere, generating a cascade of secondary particles. 
The electrons and positrons in the shower produce coherent, broadband and impulsive electromagnetic signals, that can be detected in the tens to hundreds of MHz frequency range.

Radio impulses from air-showers have been known and measured since 1960s, but it was only in the last decades, thanks to the developments in digital signal processing, that this domain experienced a rebirth as a promising astroparticle detection technique~\cite{TimsReview,Schroder:2016hrv,Connolly16}. The results of for example AERA (Auger Engineering Radio Array) and Tunka-Rex (Tunka Radio Extension) on the energy reconstruction of the primary particle~\cite{AERAEnergy,2017EPJWC.13501003S} and  the latter and LOFAR (LOw Frequency ARray) on the measurement of the mass composition~\cite{LOFARMass,Bezyazeekov:2018yjw} show that radio detection has become competitive with standard methods such as fluorescence light. 
Where LOFAR and AERA used a scintillator-based trigger, autonomous detection by a self-triggered radio detection set-up was successfully shown by TREND last year~\cite{TREND}, and earlier already by the ARIANNA~\cite{Barwick:2014pca} and ANITA~\cite{Gorham:2008yk} experiments while the AERA experiment also recorded self-triggered radio event identified as EAS signal by coincidence with the Auger Surface Detector~\cite{2012JInst...7P1023A}.
These successes are due to drastic technological advances, but also to the considerable progress made in the understanding and modeling of the radio emission mechanisms of air showers. 

Both macroscopic and microscopic approaches can be found in the literature to model the radio-emission of air-showers. The former are mostly analytical and are based on the modeling or fitting of the global physical effects that contribute to the radio emission from extensive air-showers (e.g., \cite{Kahn66,Allan:1971,Falcke03,Scholten08, Scholten18}). The latter are numerical simulations that treat particles in the air-shower individually, and compute their electromagnetic radiation from first principles. Uncertainties then stem from the calculation of the air-shower itself. Several simulation codes exist on the market with  different levels of complexity, e.g., SELFAS~\cite{SELFAS}, MGMR~\cite{Scholten08}, EVA~\cite{EVA}, CoREAS~\cite{CoREAS}, ZHAireS~\cite{Zhaires}. In the last years their results have started to converge, and were found to be consistent with radio-signal measurements taken under laboratory conditions~\cite{SLAC} and with air-shower arrays \cite{LOPES,Apel:2016gws}.

Macroscopic approaches are fast --being analytical--, and enable one to grasp a physical insight on the various features of the radio signal, that are not yet fully understood. But they also have serious limitations, when one wishes for example to study the signatures expected for specific instrument layouts. Detailed spatial, spectral and temporal structures of the signal can be easily lost in the process of integrating over different contributions and due to simplifying geometrical assumptions. Besides, these formalisms contain free parameters to be tuned, such as the drift velocities, which strongly impacts the predicted electric field strengths \cite{Scholten08}. 

On the other hand, running microscopic simulations for very high-energy particles and very large or dense arrays consisting of hundreds of antennas,  is highly time-consuming, so that one quickly reaches the limitations of the computational resources. Typically, simulating the electric-field traces over 200 radio antennas for one air-shower of primary energy $10^{17}\,\mathrm{eV}$ with the ZHAireS simulation costs about 2 hours on one node and with a thinning level of $10^{-4}$.  In the early phases of an experiment, when exploring the performances of particular layouts, a more portable and faster method is needed, that also provides as precise information than the classical models that have been studied so far. 

This can be achieved by {\it Radio Morphing}, a novel method we present in this paper.
It consists in simulating the radio signal emitted by {\it one} generic air-shower and in {\it morphing} of it in order to obtain the electric field from any primary particle, at any desired antenna position. Morphing is performed by using mostly well-documented analytical formalisms which enable to account for the effect of each relevant primary particle, as well as atmospheric conditions and detector position parameters. 

Our approach allows us to reflect the complexity of the particle distributions in spite of the analytical layer, thanks to the use of the initial full simulation output. The morphing treatment allows fast calculations. For the example given above, once a single generic shower simulation has been generated, the response over 200 radio antennas can be computed in less than 2 minutes with Radio Morphing on one node: a gain of roughly 2 orders of magnitude in CPU time. The gain further increases for lower thinning levels.

We first recall in Section~\ref{section:setup} the basics of radio-emission. Section~\ref{section:method} details the physical principles behind the construction of the Radio Morphing method, and outlines the full Radio Morphing method. We demonstrate in Section~\ref{section:validation} the performances of the method in reproducing the key features of radio-emission from air-showers, and compare our results with microscopic simulation outputs for a set of horizontal events which are expected to be measured by the future GRAND detector~\cite{Alvarez-Muniz:2018bhp}. We discuss possible improvements and limitations originated from assumptions made for simplicity in Section~\ref{section:discussion}.\\
\\
The Radio Morphing method discussed in this paper has been implemented as a dedicated Python module~\cite{RM:GitHub}, freely available online under the LGPL-3.0 license. The \texttt{radiomorphing} module requires \texttt{numpy}~\cite{numpy} in order to speed up intensive numerical computations, e.g. matrix operations or Fourier transforms.

\section{Physics of radio emission}\label{section:setup}

\begin{figure}
	\centering
		\includegraphics[width=0.45\textwidth]{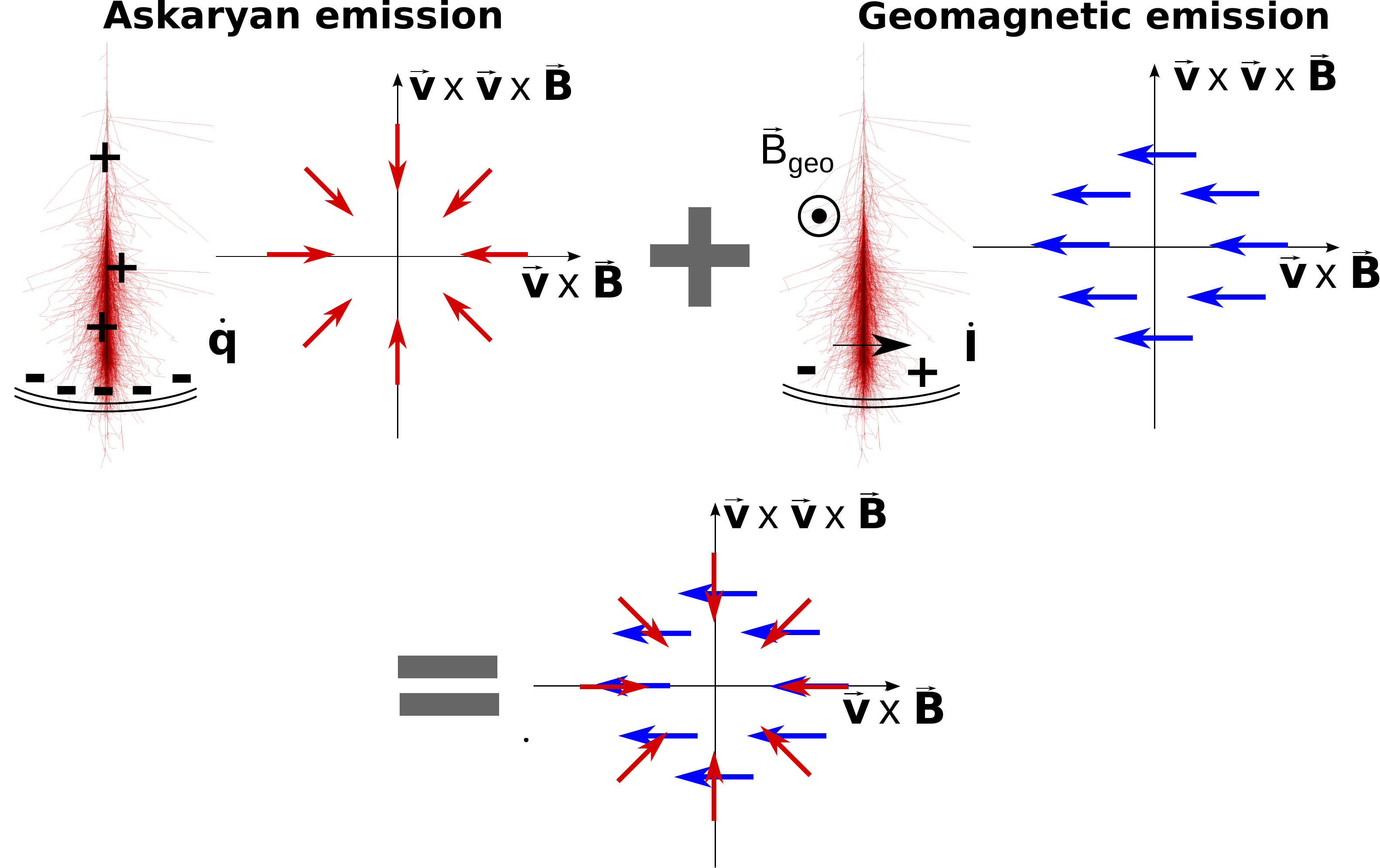}
	\caption{\label{fig:GeoAskSketch} Main radio emission mechanisms in an extensive air-shower and the polarization of their
corresponding electric field in the shower plane. The Askaryan effect can be described as a variation of the net charge excess of the shower in time ($\dot{q}$) and the geomagnetic effect results from the time-variation of the transverse current in the shower ($\dot{I}$). Both mechanisms superimpose, resulting in a complex distribution of the electric field vector (bottom).}
\end{figure}

A primary high-energy particle induces an extensive air-shower in the atmosphere of the Earth, {\it i.e.}, a cascade of high-energy, mostly leptonic, particles and electromagnetic radiation. Most of the particles are concentrated in a shower front, that is typically a few centimeter to meters thick near the shower axis.  
Time variation of the total charge or current in the relativistic shower front in combination with Cherenkov effects leads to coherent radio emission over the typical dimensions of the particle cascade. These radio pulses last typically tens of nanoseconds, with varying amplitudes of up to several hundreds of $\mu \mathrm{V/m}$. The signal can be interpreted by two main mechanisms (see figure~\ref{fig:GeoAskSketch}).

The so-called {\it Askaryan effect}~\cite{Askaryan1962,Askaryan1965} results from Compton scattering while the shower propagates through the Earth atmosphere. The resulting electrons are swept into the shower front.
In a non-absorptive, dielectric medium the number of electrons in the particle front, and therefore the net charge excess in the cascade, varies in time, which induces a coherent electromagnetic pulse. 
The radio signal emitted by this mechanism is linearly polarized. The electric field vector is oriented radially around the shower axis, so the orientation of the electric field vector depends on the location of an observer with respect to the shower axis.

The second and main mechanism at play is the {\it geomagnetic effect}~\cite{Falcke03,TimsReview}.
In the shower front, the secondary electrons and positrons are being deflected towards opposite directions by the geomagnetic field, after which they are stopped by interactions with air molecules. In total, this leads to a net drift of the electrons and positrons in opposite directions as governed by the Lorentz force $\vec{F}=q\, \vec{v} \times \vec{B}$, where $q$ is the particle charge, $\vec{v}$ the velocity vector of the shower and $\vec{B}$ the geomagnetic field vector. As these ``transverse currents'' vary in time during the air shower development, they lead to the emission of electromagnetic radiation. The polarization of this signal is linear, with the electric field vector aligned with the Lorentz force (along $\vec{v}\times \vec{B}$).

Depending on the position of the observer and hence the orientation of the electric field vector for the two emission mechanisms, these contributions can add constructively or destructively, leading to an asymmetric ring structure: a {\it radio footprint}, the pattern of the radio signal on ground.

Since the refractive index of air is slightly larger than $1$, the radio waves travel slower through the air than the relativistically moving particle front. In addition to a strong forward-beaming of the emission, this leads to a so-called Cherenkov compression. 
At particular observer positions on ground, a radio pulse is detected as being compressed in time since the radiation emitted by a significant part of the shower arrives simultaneously. The pulse becomes very narrow and coherent up to frequencies in the GHz region.

These observer positions can be found on the {\it Cherenkov ring} \cite{deVries11,Alvarez12,Nelles15}, given by $\cos\Theta_{\rm C}= (n\,\beta)^{-1}$ with $\Theta_{\rm C}$ defined as the Cherenkov angle, $n$ the refractive index of the medium that depends on the emission height, and $\beta$ the particle velocity.

One of the main observables that characterize the air-shower is the atmospheric depth $X_{\mathrm{max}}$ (also called the \textit{shower maximum} and given in $\mathrm{g}.\mathrm{cm}^{-2}$) at which the development of the cascade reaches its maximum particle number in the electromagnetic component~\cite{Xmax_plot}. Since the strength of the emitted radio signal scales linearly with the number of electrons and positrons, and since the signal is strongly beamed forward, $X_{\mathrm{max}}$ can be considered at first approximation, as a point-source position, where the maximum radiation comes from. This approximation is just valid in the far-field, and restricts therefore the methods to it.

\begin{figure*}[tb]
	\centering
		\includegraphics[width=1.00\textwidth]{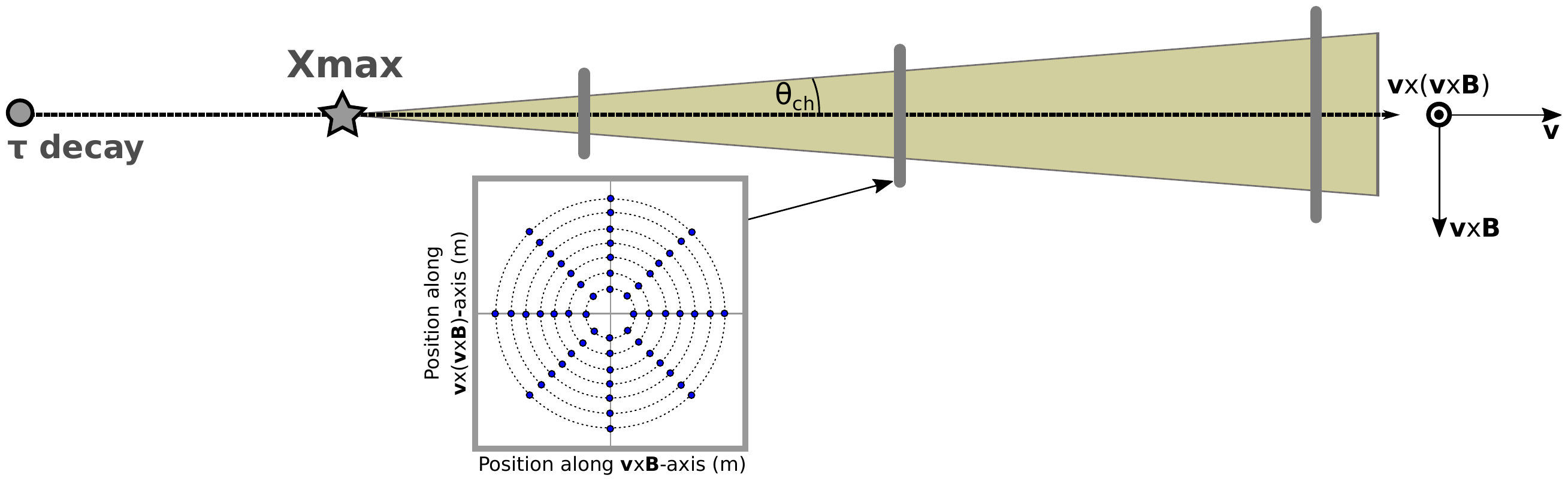}
	\caption{Sampling of the radio signal at several distances along the shower axis from the shower maximum $X_{\mathrm{max}}$. The antenna positions in the planes are arranged in a so-called star-shape pattern, defined by the shower direction $\vec{v}$ and the orientation of the Earth's magnetic field $\vec{B}$.} \label{fig:starshape_planes} 
\end{figure*}
%

\section{The Radio Morphing method}\label{section:method}

Previous macroscopic studies have shown that the average radio emission properties from air-showers depends on a limited set of parameters, describing the energy and geometry of the shower~\cite{Kahn66,Allan:1971,Falcke03,Scholten08, Scholten18}. Following this observation, we have developed the Radio Morphing method, in which radio signals are rescaled from a single reference air-shower, via a series of simple mathematical operations. 

This idea relies on the {\it universality} of the distribution of the electrons and positrons in extensive air-showers, that was pointed out by several works already~\cite{Giller05,Gora06}. Interestingly, this distribution was found to depend mainly on the depth of the shower maximum $X_{\rm max}$ and the number of particles in the cascade at that depth, that is, on the age of the shower. Based on this concept, Ref.~\cite{Universality} presented a parametrization of the air-shower pair distributions, that enables one to calculate the properties of any air-shower by a linear rescaling of a small number of parameters. The parametrization was later refined by, e.g., Refs.~\cite{Giller15,Smialkowski18}.

The distribution of these electrons and positrons in the shower front are directly responsible for the radio emission. Hence the associated radio signals are also expected to be universal. 
Because the radio signal is integrated over the full shower evolution, shower-to-shower fluctuations are further smoothed out for observer positions in the far-field and this average universality can be seen as a robust estimate. 
Here, we would like to mention that for energies of electromagnetic showers of about $10^{20}\,$eV the Landau-Migdal-Pomeranchuk (LPM) effect which leads to a suppression of bremsstrahlung and pair production has to be taken into account~\cite{Cillis:1998hf}. Below the energy, the effect will not produce serious distortion in the shower development.

The strength of the measured radio emission is impacted by other external ingredients such as the geomagnetic angle, and various geometrical distance scales (shower zenith angle and altitude) that can modify the distance of the observer to the radio source and thus stretch the size of the radio footprint on ground. First we focus on the generation of a reference shower in Section~\ref{section:observers}. We will demonstrate in Section~\ref{section:amplitude} that, taking these mentioned processes into account, the radio emission properties can still be parametrized by only 4 parameters with a good level of precision at a fixed distance from the shower maximum $X_{\rm max}$, namely the primary particle energy ${\cal E}$, the shower zenith and azimuth angles $\theta$ and $\phi$ at injection, and the shower injection altitude $h$. The direction and position of the shower can then be re-adjusted using isometries (Section~\ref{section:isometries}). 
The last step is the performance of two-point interpolations (Section~\ref{sec:interpolation}) of the electric field trace at the desired antenna position.
Hence, we are able to calculate, at any desired observer positions, the electric field $\vec{E}(\vec{x},t)$ emitted by any target shower from one generic simulated shower, acting as reference, by simple analytical operations.

\begin{figure}[t]
	\centering
		\includegraphics[width=0.45\textwidth]{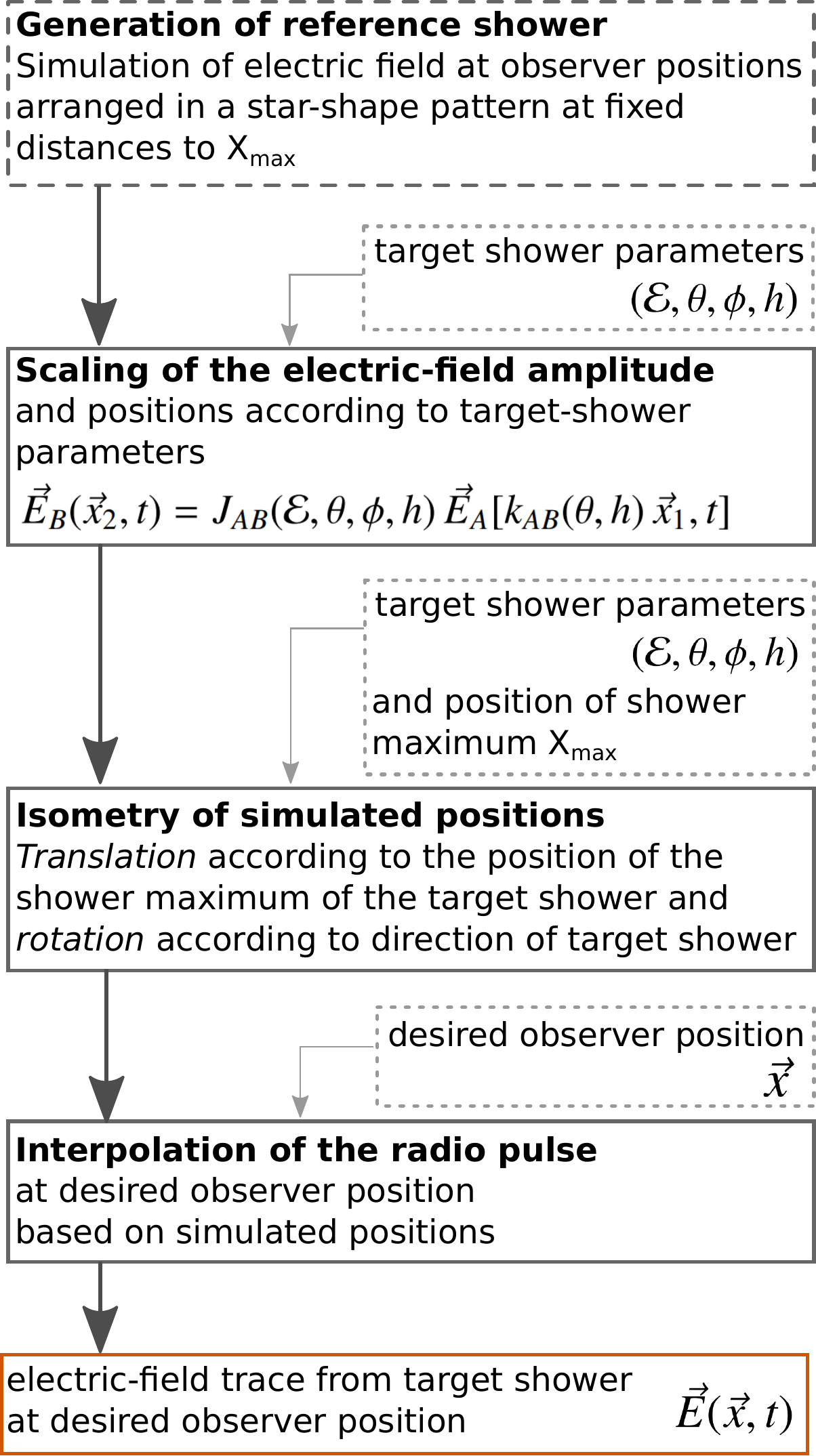}
	\caption{Recipe for Radio Morphing: These different steps are applied to a reference shower to receive the electric field traces for a target shower with desired parameters at the desired observer positions.}
\label{fig:Recipe}
\end{figure}

The different steps of the Radio Morphing method are summarized in Fig.~\ref{fig:Recipe}. The corresponding code is publicly available, see \cite{RM:GitHub}.

\subsection{Sampling positions for the simulated radio signal}\label{section:observers}
In the simulation of radio emission from air showers the signals are recorded at a set of observer positions.
To select these positions wisely, we profit from our knowledge about the emission mechanisms:

As mentioned in the previous section, the geomagnetic and the Askaryan effects are linearly polarized along $\vec{v}\times\vec{B}$ and around the shower axis, respectively. This naturally leads to a radiation profile that is not rotationally symmetric, and that can be adequately described in the shower-coordinate system defined by $(\vec{v}, \vec{v}\times\vec{B}, \vec{v}\times\vec{v}\times\vec{B})$. The advantage of the shower coordinates is that the radio emission can be fully described by a superposition the two main emission mechanisms whose contributions can be disentangled due to their different polarizations.

A correct sampling of the radio signals has to record the lateral distribution of the radio signal as well as the longitudinal distribution function along the shower axis defined by $\vec{v}$.
In order to guarantee it while minimizing the number of simulated observer positions, the positions are optimized so that at several distances from $X_{\rm max}$ along the shower axis they cover the locations where the interference between the two radiation components reach their minima and maxima. 

The variations in signal strength are along the $\vec{v}\times\vec{B}$-axis. We thus position a set of observers over a {\it star-shaped pattern} in the shower plane with eight arms, two aligned with the $\vec{v}\times\vec{B}$ axis, and two aligned with the $\vec{v}\times\vec{v}\times\vec{B}$-axis (see figure~\ref{fig:starshape_planes}) \cite{Buitink14}. 
The extensions of these star-shape pattern in several distances from $X_{\rm max}$ are determined by the fact that the emission is strongly beamed forward and forms a cone of a few degree opening angle with $X_{\mathrm{max}}$ as an approximation for a point source (see Fig.~\ref{fig:starshape_planes}).

In the following, reference showers are produced via microscopic simulations of the radio signal, performed in Cartesian coordinates, with the $x$-component aligned along magnetic North-South (NS), the $y$-component along East-West (EW) and the $z$-component pointing upward (Up), while the scaling procedure is performed in the shower referential defined by ($\vec{v}$,$\vec{v}\times\vec{B}$,$\vec{v}\times\vec{v}\times\vec{B}$).

\subsection{Parametrizing the electric field}\label{section:amplitude}

The radio signal measured at a given position $\vec{x}$ and time $t$ can be formally described by the electric field vector $\vec{E}(\vec{x},t)$. Four parameters constrain the strength and polarization of the field at a given observer position: the primary's energy ${\cal E}$, the shower direction towards which the cascade is propagating\footnote{In the conventional definition, azimuth and zenith are defined to describe the direction from which the shower is coming.}, defined by zenith $\theta$ and azimuth $\phi$ at injection, and the altitude of injection $h$.
The key hypothesis in Radio Morphing is that, at a fixed distance to the shower maximum $X_{\rm max}$, 
the electric field vector of any shower $B$ at the position $\vec{x}_B$ can be derived from that of a reference shower $A$ by a set of simple operations that are applied on the overall electric field  $\vec{E}_A$ and on the position $\vec{x}_A$
\begin{equation}
\vec{E}_B(\vec{x}_B,t) = J_{AB}({{\cal E},\theta,\phi,h}) \,\vec{E}_A[k_{AB}({\theta,h})\,\vec{x}_A,t] \ ,
\end{equation}
where the scaling matrices $J_{AB}({{\cal E},\theta,\phi,h})$ and factors $k_{AB}({\theta,h})$, can be calculated as a function of the reference and target shower parameters $({\cal E},\theta,\phi,h)$, taking into account independent effects related to the primary energy, the geomagnetic field, the air density, and air refraction index. We detail in this section the dependency of the electric field on these various physical parameters at play in order to express $J_{AB}$ and $k_{AB}$. 

We will derive their mathematical formulae by expressing their dependency on the primary energy ${\cal E}$, the geomagnetic angle $\alpha$, the density $\rho$ at the height of the shower maximum and the corresponding values for the Cherenkov angles $\Theta_{\rm C}$ of the target and the reference showers: 
\begin{equation}
J_{AB}({{\cal E},\theta,\phi,h}) = j_{\cal E}  j_{\rho} j_{\rm C} J_\alpha 
\end{equation}
and 
\begin{equation}
k_{AB}({\theta,h})= 1/j_{\rm C},
\end{equation}
where the scalars $j_{\cal E}$, $j_{\rho}$, $j_{\rm C}$ and the matrix $J_\alpha$ will be derived or explained below.
The electric-field and position vectors will be written in the shower coordinate system $(\vec{v}, \vec{v}\times\vec{B}, \vec{v}\times\vec{v}\times\vec{B})$.


\subsubsection{Scaling in primary energy}\label{section:energy}

Up to the frequencies corresponding to the typical thickness of the particle shower front (a few meters), secondary particles in the air-shower create a coherent radio emission: at a given frequency, the radiation emitted from several particle experiences negligible relative phase shifts during its propagation to the observer. This implies that the vectorial electric fields produced by each particle also add up coherently, and the total electric field amplitude scales linearly with the number of particles, itself scaling with ${\cal E}$. We can thus write $|\vec{E}(\vec{x},t)| \propto {\cal E}$. 

This relationship is consistent at first order with the seminal expression of the electric field amplitude derived by \cite{Allan:1971} from pioneering measurements, and with the recent measurements performed by AERA and LOFAR \cite{AERAEnergy,LOFARMass}.

In practice, the electric-field amplitude of a generic shower $A$ with primary energy ${\cal E}_A$ can be scaled to the one of a target shower $B$ with the energy ${\cal E}_B$, by multiplying by the factor\begin{equation}
j_{\cal E} = \frac{{\cal E}_B}{{\cal E}_A}\ .
\end{equation}

Note that for frequencies typically higher than $\sim100\,$MHz, coherence in the emission is no longer guaranteed (besides for position in the Cherenkov cone, see Sec.~\ref{section:setup}) and can lead to uncertainties while applying this factor. More precisely, the effective thickness of the shower front that sets the coherence condition also depends on the observer angle.

\subsubsection{Scaling in geomagnetic angle}\label{section:geomagnetic}

The strength of the radiation emitted via the geomagnetic effect scales with the strength of the Lorentz force $\vec{F}=q \, \vec{v} \times \vec{B}$ experienced by each particle in the shower front, and that induces a transverse current. The magnitude of the emitted signal scales with  
$|\vec{v}\times \vec{B}|=|\vec{v}| |\vec{B} | \sin\alpha$,
leading directly to a sinusoidal dependency of the electric field strength over the geomagnetic angle: 
$|\vec{E}| \propto |\vec{v}\times \vec{B}| \propto  \sin\alpha$. 
Here, the geomagnetic angle is given by $\alpha= \measuredangle (\vec v, \vec B)$, introducing the dependency on the angles $\theta$ and $\phi$ of the shower.
Also, this dependency is consistent at first order with the seminal expression of the electric field amplitude derived by \cite{Allan:1971} from pioneering measurements, and is confirmed experimentally by recent measurements performed by CODALEMA \cite{Ardouin:2009zp} and later by AERA and LOFAR \cite{AERAEnergy,LOFARMass}. 
We neglect the linear dependency on the local magnetic field strength at the moment and choose a reference shower which is simulated for the target site.

The amplitudes of reference shower $A$ and target shower $B$ can be related via a scaling factor that takes into account the two geomagnetic angles $\alpha(\theta_A,\phi_A)$ and $\alpha(\theta_B,\phi_B)$ sensed by each shower: $j_\alpha=\sin\alpha(\theta_B,\phi_B)/\sin\alpha(\theta_A,\phi_A)$. This scaling was recently demonstrated experimentally by AERA \cite{AERAEnergy}.
	
Since the geomagnetic emission is linearly polarized along $\vec{v}\times \vec{B}$, this factor is multiplied to the ${\vec{v}\times\vec{B}}$ component of the electric field in shower coordinates. 
Thus the scaling matrix can be expressed in the shower referential by
\begin{equation}
J_\alpha = 
\begin{bmatrix}
1 & 0 & 0 \\
0 & j_\alpha & 0 \\
0 & 0 & 1 \\
\end{bmatrix}
\quad \mathrm{with} \quad 
j_\alpha=\frac{\sin\alpha(\theta_B,\phi_B)}{\sin\alpha(\theta_A,\phi_A)}
\ .
\end{equation}

Here, we neglect the possible projection of the geomagnetic component into the $\vec{v}$ component of the electric field, assuming that it is negligible with respect to our target accuracy. The phase shift between Askayran and Geomagnetic emission, as observed by LOFAR~\cite{Scholten:2016gmj}, is included implicitly by using a microscopic simulation as reference. We assume that the phase shift depends solely on the observed frequency and on the position of the observer with respect to the Cherenkov cone. Therefore, a scaling of the amplitude will not change the time difference, determined by the microscopic simulation.

For simplicity, we chose in the current version of Radio Morphing to scale ${\vec{v}\times\vec{B}}$ component of the electric field without an a-priori decoupling of the contributions from the Geomagnetic and Askaryan effect. This impact of the latter one could be of order of a few$\,\%$ contribution, depending on the measured polarization~\cite{14percent}.
To exclude this artificially induced uncertainty, one has to decouple the two emission components completely, e.g. by comparing simulations with magnetic field turned on and off. These effects, as well as a possible scaling with the magnetic field strength, will be included in a future version of the Radio Morphing code.

\subsubsection{Effects of air density}\label{sec:height}
The parametrization on the injection altitude and zenith angles, and therefore on the air density, is poorly documented in the literature. Complicated ad-hoc fits have been invoked \cite{Glaser16,Scholten18}, and the zenith scaling is handled at the moment in the community as a ``distance to $X_{\rm max}$'' correction. We present here a more natural modeling of these effects, validated with ZHAireS simulations.

In practice, our input parameter is the initial injection height $h$ of the air shower, that corresponds to the altitude at which the shower starts developing in the atmosphere. Geometrically, this altitude is connected to the altitude of the shower maximum $h_{X_\mathrm{max}}$ by the following relation via the zenith angle of the air shower $\theta$ at injection:
\begin{equation}\label{eq:height}
		h_{X_{\mathrm{max}}} = h + d_\mathrm{hor}/\tan \theta
\end{equation}
with $d_\mathrm{hor}$ the horizontal distance of the shower maximum to the injection point of the shower. Here, a flat-Earth approximation can be used since $d_\mathrm{hor} \ll R_\otimes$.

\begin{figure}[tb] 
	\centering
		\includegraphics[width=0.45\textwidth]{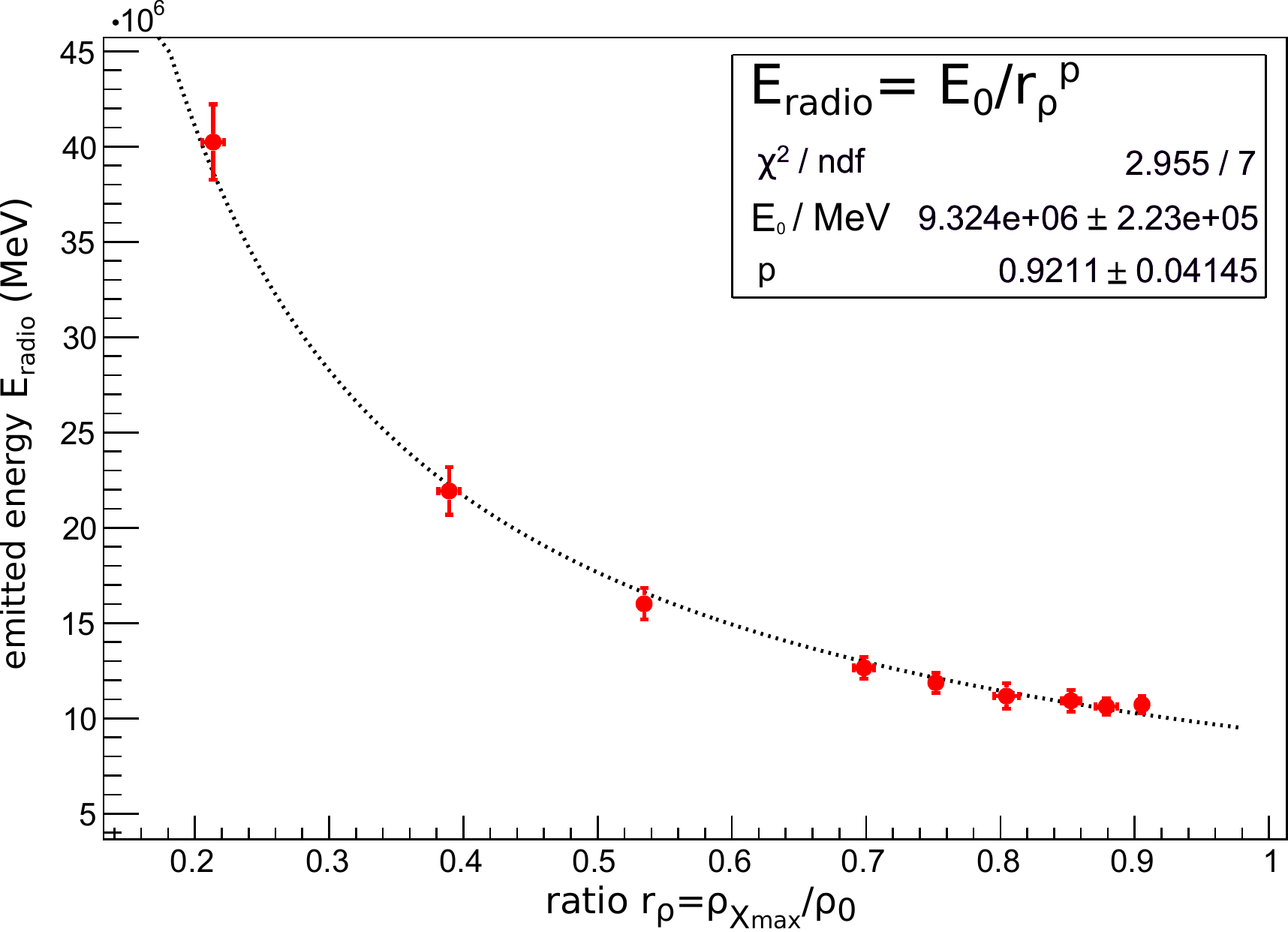}\\
	\caption{Density scaling of the radio signal in the $30-80\,\mathrm{MHz}$ frequency band for a proton-induced showers with energy $10^{17}\,$eV arriving from the North.
Red dots show the simulation outputs for an antenna array at the AERA site: the radiated energy in the radio signal $E_\mathrm{radio}$ as a function of the density ratio $r_\rho=\rho_{X_{\rm max}} / \rho_0$, with $\rho_{X_{\rm max}}$ the actual density at the height of the shower maximum and $\rho_0= 1.225\,\mathrm{g/cm}^3$ the density at sea level. The data points can be well-fit by a power-law function $E_\mathrm{radio}=E_0/{r_{\rho}}^p$ where the values of $E_0$ and $p$ are indicated in the label (dotted line).}
	\label{fig:density_scaling} 
\end{figure}

The dependency of the radiated energy on the density at $X_{\rm max}$ is complicated to estimate analytically, as it corresponds to integrated values over the full shower development. We investigated this effect numerically using ZHAireS \cite{Zhaires} simulations. We have produced sets of shower simulations with the parameters of the AERA \cite{AERAEnergy} site at the Pierre Auger Observatory. 
The sets of simulations contain proton-induced showers with energy $10^{17}\,$eV arriving from the North and a zenith angle $\theta = 80^\circ$ in the conventional angular system used for cosmic rays. To achieve different densities at the position of the shower maximum, we changed the injection height of the shower artificially. We calculated for each event the measured radiated energy on ground emitted by the shower.
The result is illustrated in Figure~\ref{fig:density_scaling}. The errorbars are given by the RMS for 10 simulated air showers for each shower geometry.
Interestingly, we find that the scaling of the radiated energy scales as $|\vec{E}|^2 \propto [\rho_{X_{\rm max}}(h,\theta)]^{-1}$ (for simplicity we use a value of $1$ instead the exact fitting result), where $\rho_{X_{\rm max}}(h,\theta)$ is the air density at $X_{\rm max}$, related to the density at $h$ via Eq.~\ref{eq:height}. This effect might be understood as follows.
Each part of the footprint on the ground is more sensitive to a specific part of the longitudinal profile of the shower $X$. Thus if one ``separates'' each part of the footprint and links it to a specific position $X$ in the profile, the electric field will scale with the air density at $X$. The integration over the energy in radio that reaches the ground, brings up an intricate convolution on the number of particles at each particular depth $X$, the air density at that depth and the position on the ground (each part is more sensitive to a certain $X$). Nevertheless, this effect is not yet fully understood and will be investigated further in a future study.

Even though we expect that the charge-excess increases with the air density, this effect is found to be relatively small. Therefore, we neglect it for the moment. In addition, the Askaryan emission is just a minor correction to the total measured peak amplitude for inclined showers. For the reference shower used in Sec.~\ref{sec:statvalid},we found a relative contribution to the peak amplitude on the Cherenkov cone of about $14\%$. Therefore, we apply the scaling to all components of the electric field vector in the current version of Radio Morphing.

From these results, we can compute the scaling factor to obtain the amplitude of the target shower $B$ from the reference shower $A$ 
\begin{equation}
j_{\rho}= \left[\frac{\rho_{X_{\rm max}}(h_A,\theta_A)}{\rho_{X_{\rm max}}(h_B,\theta_B)} \right]^{1/2}
\end{equation}
with $\rho_{X_{\rm max}}(h,\theta)$ the air density at altitude $X_{\rm max}$ for reference and target showers $A$ and $B$, related to the altitude at shower injection via Eq.~\ref{eq:height}.
A comparison with the formula presented in Ref.~\cite{Glaser16} shows a nice agreement between the two formalisms, for highly inclined showers with high densities at the shower maximum.

\subsubsection{Stretching effect from the Cherenkov angle}\label{sec:cherenkov}

The injection altitude and zenith angle also impact the size of the radio footprint. 
The opening angle of this cone depends on the altitude as the atmosphere becomes denser with decreasing height, leading to a larger refractive index, and hence a larger Cherenkov cone. Indeed, the radius of the Cherenkov cone within which the radio signal is emitted is given by $r_L=L \tan \Theta_C$ where $L$ is the distance from $X_{\mathrm{max}}$ and $\Theta_C =\arccos[1/n(h)]$ is the Cherenkov angle \cite{deVries11,Alvarez12}. The refractive index $n_{X_{\rm max}}(h,\theta)$ is a function of the altitude $h$ and zenith angle $\theta$ and has to be evaluated at the altitude of the shower maximum $X_{\mathrm{max}}$ using Eq.~\ref{eq:height}. For instance in ZHAireS, an exponential function is implemented~\cite{Zhaires}.

The scaling between two showers at different injection heights is given by the ratio between the Cherenkov radii. This leads to a stretching factor for the radio footprints, to be applied to the positions $\vec{x}_B = k_{\rm C} \vec{x}_A$ with 
\begin{equation}
k_{\rm C} = \frac{\Theta_{{\rm C},B}}{\Theta_{{\rm C},A}} \sim \frac{\arccos[1/n_{X_{\rm max}}(h_B,\theta_B)]}{\arccos[1/n_{X_{\rm max}}(h_A,\theta_A)]}
\end{equation}
with $(h_A,\theta_A)$ and $(h_B,\theta_B)$ the parameters of reference and target showers $A$ and $B$. Here we have assumed that $\tan\Theta_{\rm C}\sim \Theta_{\rm C}$, as $\Theta_{\rm C}\ll 1$. This means that the distances between the simulated antenna positions along the star-shaped arms are corrected by the corresponding stretching factor $k_{\rm C}$.

By energy conservation, the total radiated energy over each area intersecting the Cherenkov cone should be constant: $\lvert \vec{E}(\vec{x},t) \rvert ^2/r^2 = {\rm constant}$, where $r$ is the radius of the intersected area. The stretching of the area by $r_B = k_{\rm C} r_A$ yields the following scaling factor between showers $A$ and $B$
\begin{equation}
j_{\rm C} = \frac{\lvert \vec{E}_B(\vec{x},t) \rvert}{\lvert \vec{E}_A(\vec{x},t) \rvert}  = \frac{1}{k_{\rm C}} \ .
\end{equation}

Note that very deep showers can be affected by the clipping effect, when particles reach the ground before the shower has completely developed. This effect is not accounted for in the scaling of the electric-field amplitude.

\begin{figure}[tp]
	\centering
		\includegraphics[width=0.45\textwidth]{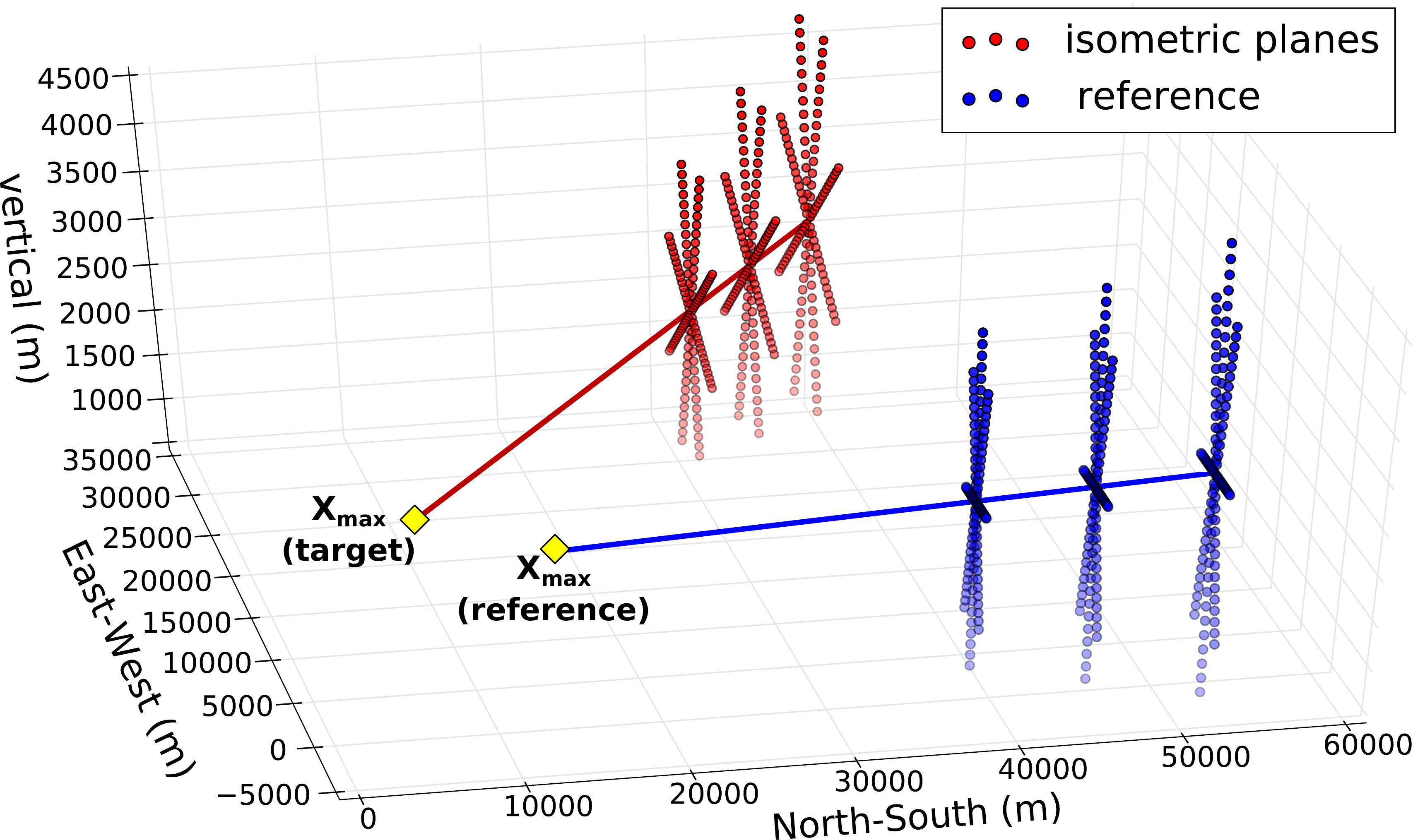}
	\caption{\label{fig:scaling_positions} Illustration of the isometry operation: the simulated antenna positions for the reference shower (blue) are rotated and translated accordingly to the new shower direction (red line) and the $X_\mathrm{max}$ position (yellow diamonds) of the target shower (red).}
\end{figure}

\subsection{Isometries of observer positions} \label{section:isometries}
Once the electric field vector of the reference shower $\vec{E}_A(\vec{x}_1,t)$ has been morphed to $\vec{E}_B(\vec{x}_2,t)$ according to the set of parameters $({\cal E},\theta,\phi,h)$ of the target shower, we rotate and translate the positions, at which the signal was simulated, according to the new shower direction. This can be done straightforwardly by a rotation and a translation of observer positions in the star-shaped planes, where the electric field of the reference shower were sampled (see Section~\ref{section:observers}), as is illustrated in Fig.~\ref{fig:scaling_positions}.

The isometries performed on the $\vec{E}_A(\vec{x}_A,t)$ should (by definition) conserve the distance of each star-shaped plane to the position of $X_{\rm max}$. This condition is required in order to ensure the validity of the morphing process performed in the first step to account for the parameters $({\cal E},\theta,\phi,h)$.

The physical location in space of the target's shower maximum $X_{\rm max}$ depends on the actual shower parameters, as e.g. the primary's energy, as well as on the primary-particle type. It can be obtained from e.g. dedicated simulations of the induced particle cascade or be computed as follows: one integrates the traversed air density along the shower axis from the point of shower injection until the average depth of the shower maximum for the target shower is reached for this specific primary type, taking into account an atmosphere density model (e.g., an isothermal model).

\begin{figure}[t]
	\centering
		\includegraphics[width=0.45\textwidth]{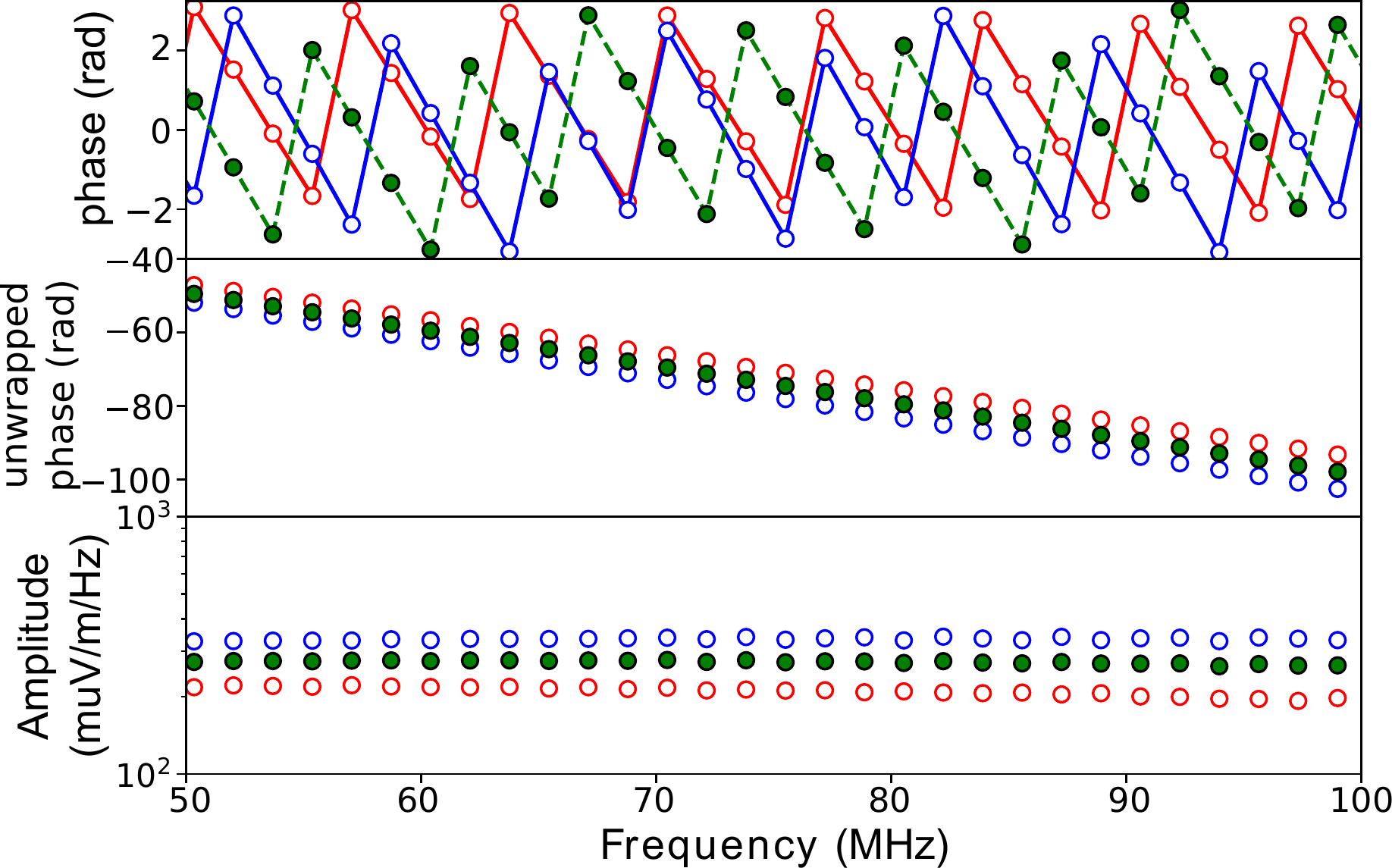}
	\caption{\label{fig:interpolation} Example for the interpolation of phase (wrapped and unwrapped) and amplitude for one antenna position: the red and blue markers represent the values at the antenna positions $a$ and $b$ which are used for the interpolation of the signal at the desired positions. Resulting phase and amplitude are represented by the green markers.}
\end{figure}

\subsection{Interpolation of the electric field traces: $\vec{E}_A(x_i,t) \rightarrow \vec{E}_A(x,t)$}\label{sec:interpolation}

Once the electric field has been sampled at fixed antenna positions in the star-shaped planes, we interpolate the signal pulses at any desired position. 

In the provided scripts for Radio Morphing, we implemented the method presented in \cite{EwaThesis} as an example for an interpolation of radio signals. Here, a Fourier transform into the frequency domain is performed, and the signal at a given location is obtained by a linear interpolation in the frequency domain between two generic positions.

The signal spectrum can be represented in polar coordinates as
	\[f(r,\varphi)=r e^{i\varphi}	\]
with $r$ as the signal amplitude and $\varphi$ the complex phase in the interval $\left[-\pi, \pi\right)$. We detail in the following the two-point interpolation of $r$ and $\varphi$ at a given observer position. 

The spectrum phase is wrapped into the interval $\left[-\pi, \pi\right)$, which results in discontinuities at the interval limits, and to sawtooth-shaped features (see Fig.~\ref{fig:interpolation}). Phases have to be unwrapped before interpolation, in order to account for data points sharing the same frequencies but with shifted phases that are not on the same saw-tooth edge. To locate the discontinuities the following conditions with $i=2,...,n$ are scanned:
\begin{eqnarray}
	\left| \varphi_i -\varphi_{i-1} \right| &>& \left|  \varphi_i -\varphi_{i-1} \right| + \pi \quad \mbox{ for } \varphi_i <\varphi_{i-1} \\
	 \left| \varphi_i -\varphi_{i-1} \right| &<& \left|  \varphi_i -\varphi_{i-1} \right| - \pi \quad\mbox{ for } \varphi_i >\varphi_{i-1} \ .
\end{eqnarray}

For each discontinuity fulfilling these conditions, all the following data points $\varphi_{i+m}$ are then corrected for a constant additive with $l$ as the number of preceding discontinuities:
\begin{equation}
\varphi_{i+m, {\rm new}}= \varphi_{i+m} +2\pi  l \ .
\end{equation}
This algorithm requires a sufficient sampling rate of the spectrum since the phase difference between consecutive data points has to be smaller than $\pi$. 
The unwrapping results in continuous phases can be interpolated linearly:
\begin{equation}\label{eq:phase}
	\varphi(\vec{x})= c_a  \varphi(\vec{x}_a) + c_b  \varphi(\vec{x}_b)\ .
\end{equation}
	Here, $\vec{x}$ is the observer position of the interpolated signal, and $\vec{x}_a$ and $\vec{x}_b$ the actual simulated observer positions. The weighting coefficients $c_a$ and $c_b$ are defined as:
	
		\[ c_a= \frac{\left|\vec{x}_a-\vec{x}\right|}{\left|\vec{x}_b-\vec{x}_a\right|} \quad \mathrm{ and } \quad c_b= \frac{\left|\vec{x}_b-\vec{x}\right|}{\left|\vec{x}_b-\vec{x}_a\right|}
	\]

Since the amplitude in the frequency domain is independent in each frequency bin, a linear interpolation within one bin is sufficient for $r$:
\begin{equation}
r(\vec{x})= c_a  r(\vec{x}_a) + c_b  r(\vec{x}_b)\ .
\end{equation}
	
The interpolated spectrum is then given by
\begin{equation}
f_{\rm int}(r,\varphi)=r(\vec{x}) \,e^{i\varphi(\vec{x})}
\end{equation}
from which the corresponding time series can be derived via inverse Fourier transformation. An example for the interpolation of the phase and the amplitude is shown in figure~\ref{fig:interpolation}.

%
The linear interpolation of the phases (see Eq.~\ref{eq:phase}) implies a linear interpolation of the arrival time as long as the wave front can be estimated as a plane between two simulated observer positions, which is valid in the case of a dense grid of observers. This is a simplification of the hyperbolic shape of the wave front, that holds if the distance between the antennas is in the order of the wavelengths.

In the example presented in Section~\ref{section:validation}, the distances between the simulated antenna positions of the reference shower are larger than the radio wavelengths considered. In that case, the phase gradient in the phase interpolation cannot reproduce correctly the arrival timing of the signal at the considered antenna positions, while the signal structure itself is not affected.  The correction of the arrival time of the signal at the observer position is part of foreseen developments for the Radio Morphing method.

\begin{figure}[t]
	\centering
		\includegraphics[width=0.48\textwidth]{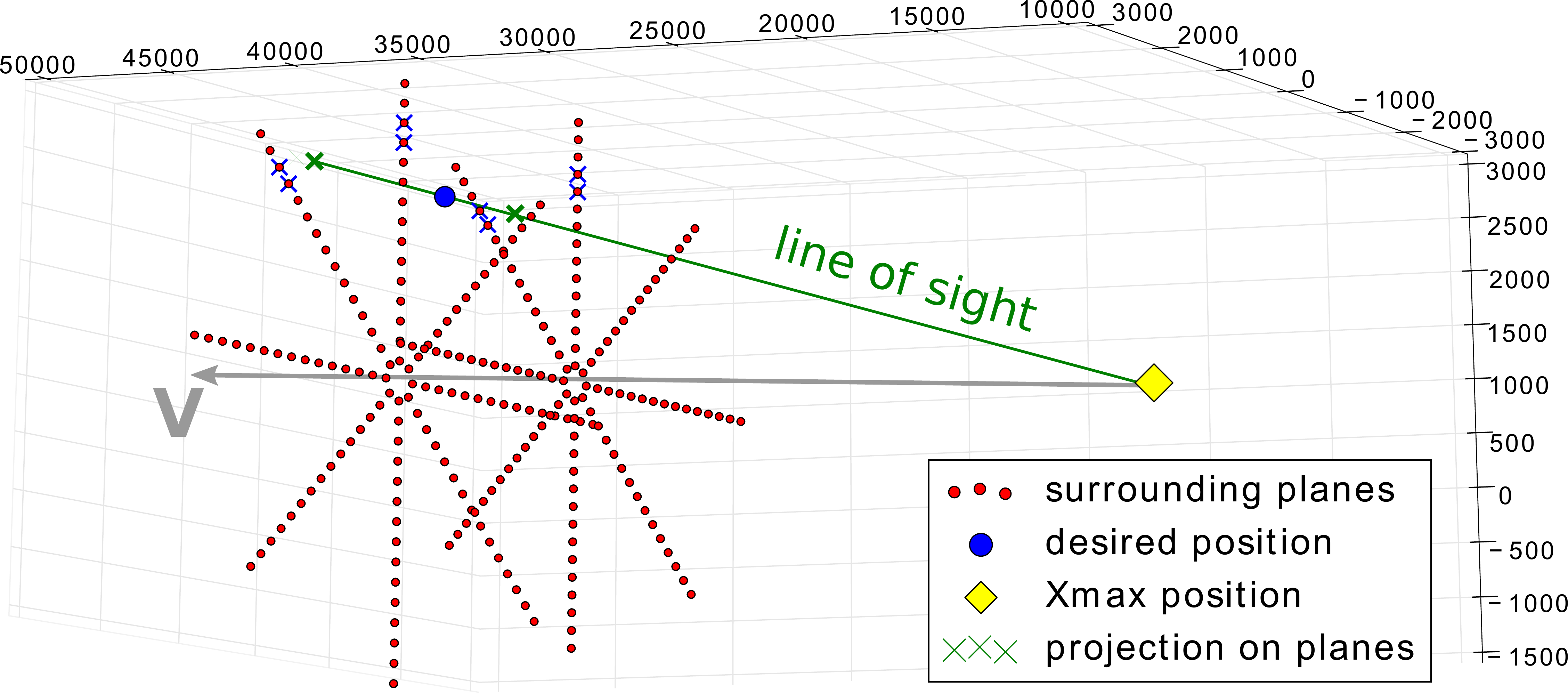}
	\caption{\label{fig:3Dinterpolation_new} Example for interpolation in 3D: after finding the two closest star-shape planes (red dots) to the desired antenna positions (blue circle), the projection of this position onto the planes along the line of sight are determined (green crosses). The 4 closest neighbored antenna positions which were simulated get identified (blue crosses). The yellow diamond represents the position of $X_{{\rm max}}$ for that example shower.}
\end{figure}

\subsubsection{3D interpolation}

The electric field time trace can be computed at any location inside the conical volume resulting from the isometry transformation. The process is the following, also illustrated on Fig. ~\ref{fig:3Dinterpolation_new}: 
\begin{itemize}
\item First we define the intersection between the line linking the observer to the $X_{\mathrm{max}}$ position and the two star-shape planes surrounding the observer position. 
\item Then we compute the signal for each of these two intersection points from the closest neighbor's signals through a bilinear interpolation, using the method  detailed above. 
\item Finally, we interpolate these two signals in order to compute the time trace at the desired observer position. 
\end{itemize}
The underlying hypothesis of this treatment is that the radio emission is point-like, and emitted from $X_{\mathrm{max}}$.

\begin{figure}[tb]
	\centering
		\includegraphics[width=0.45\textwidth]{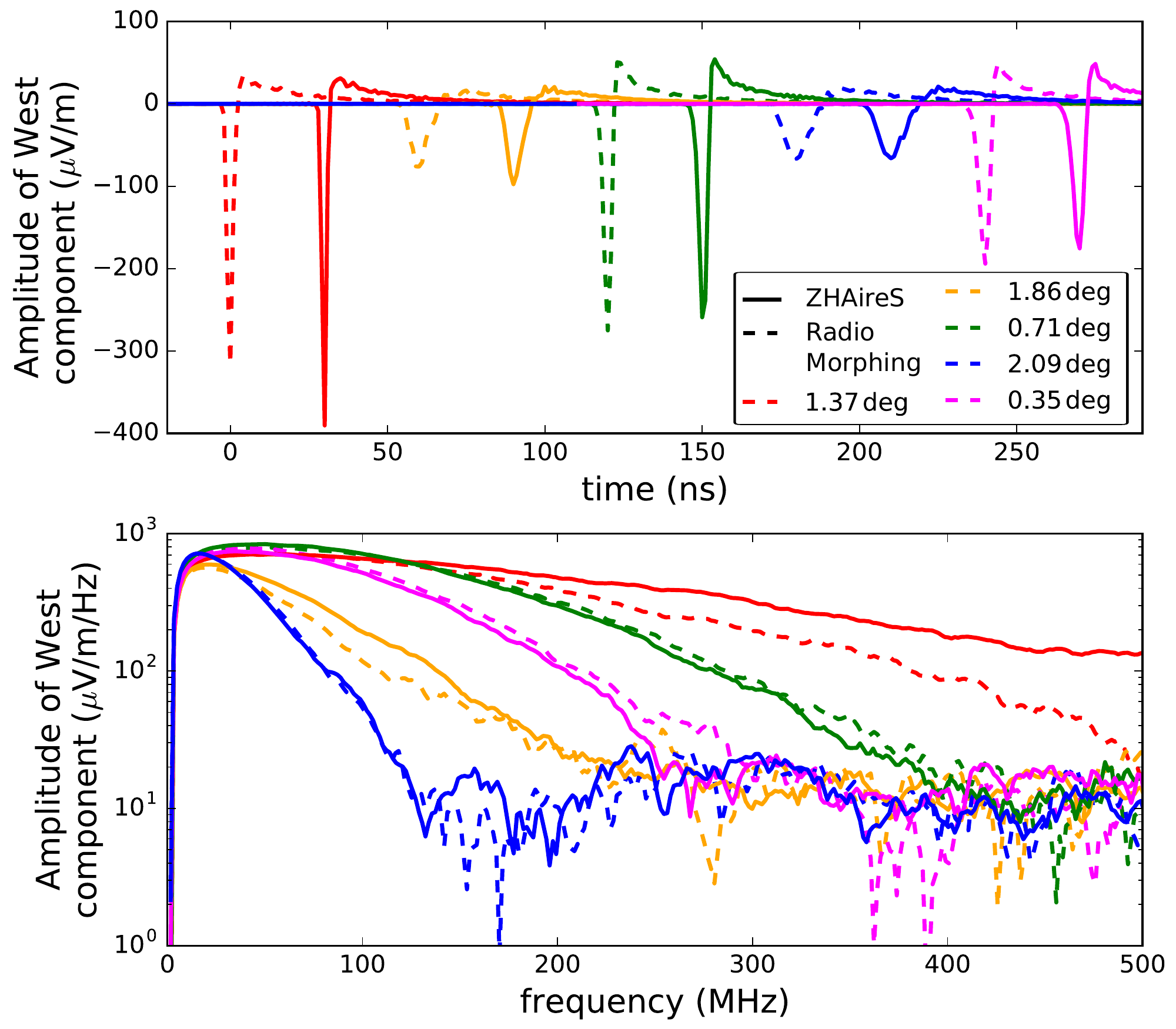}
	\caption{\label{fig:timetraces_spectra} Example signal traces for an air shower induced by an electron with an energy of $1.05\,$EeV, an azimuth angle of $50^\circ$ and a zenith angle of $89.5^\circ$ (slightly up-going shower) using Radio Morphing (solid) and ZHAireS (dashed lines) for comparison. The antenna positions are in $\sim 75\,$km distance along the shower axis to the shower maximum while the Cherenkov ring is expected to be at an off-axis angle of $\sim 1.4^\circ$. Top: Time traces of the West-East component of the electric field for antenna positions in different distances to the shower axis, given as off-axis angle, in the time domain. The time traces were shifted in time for a better comparison. Bottom: The corresponding frequency spectra.}
\end{figure}

\section{Comparison to microscopic simulation and performances}\label{section:validation}

\subsection{Time traces and frequency spectra}
We validate the behavior of Radio-Morphed time traces and frequency spectra by a direct comparison to microscopic simulations for various antenna positions.
As a reference shower we used an electron-induced air-shower with a primary energy of $0.1\,$EeV, an azimuth angle of $230^\circ$ and a zenith angle of $88.5^\circ$ (slightly up-going shower), injected at a height of $1700\,$m above sea-level. The radio emission was sampled in $5\,$km steps at several distances from the shower maximum. We simulated the signal for $184$ observer positions per plane arranged in the star-shape pattern, so that a conical volume with a half-angle equal to 3$^{\circ}$ is covered. A thinning level of $10^{-4}$ and a sampling rate of $10\,$GHz were set for the simulation. The altitude is $2000\,$m above sea-level. The magnetic field direction is describe an inclination of $63.18^\circ$ and a declination of $2.72^\circ$, having a magnetic field strength of $56.5\,\mu$T. As an atmosphere, we are using the Linsley model including coefficients for the US standard atmosphere. These are the default options for all following ZHAireS simulations.

Figure~\ref{fig:timetraces_spectra} shows  radio signals associated with a ZHAireS simulation (dashed lines) and Radio Morphing computation (solid lines) of an example target shower induced by an electron with energy of $1.05\,$EeV, first interaction at an height of $2200\,$m above sea-level, azimuth angle of $50^\circ$ (propagating towards North-West)  and zenith angle of $89.5^\circ$ (slightly up-going shower). For the ZHAireS simulation a thinning level of $10^{-4}$ was set.
The signal is calculated for positions at several distances to the shower axis which are defined as an off-axis angle, located at a distance of roughly $d=75\,$km from the injection point of the shower.
The bipolar characteristic of the signals is correctly reproduced in the time domain, and the signal amplitudes agree well with each other.

One can see that the time-compression of the signal at the Cherenkov angle is also preserved by Radio Morphing. The Cherenkov ring is estimated to be located at $\sim 1.4^\circ$ from the shower axis, although this value strongly depends on the model used to calculate the refractive index $n_{X_{\rm max}}$ at the altitude of $X_\mathrm{max}$. 
As expected, the time traces and the frequency spectra both increase in amplitude at high frequencies when approaching the Cherenkov ring (red lines). 
A slight mismatch in signal amplitudes is observed for the position closest to the Cerenkov ring. This is very certainly due to statistical fluctuations in the shower development, which induce a difference between the $X_{\mathrm{max}}$ value of the simulated shower compared to that computed with Radio Morphing (see Section~\ref{section:isometries}). These different $X_{\mathrm{max}}$ values induce a (purely geometrical) variation of the Cherenkov ring radius at ground, only partially compensated by the different values of the Cherenkov angle at different $X_{\mathrm{max}}$ heights.

\subsection{Peak amplitude distributions}\label{sec:comp1}

\begin{figure}[tb] 
	\centering
		\includegraphics[width=0.45\textwidth]{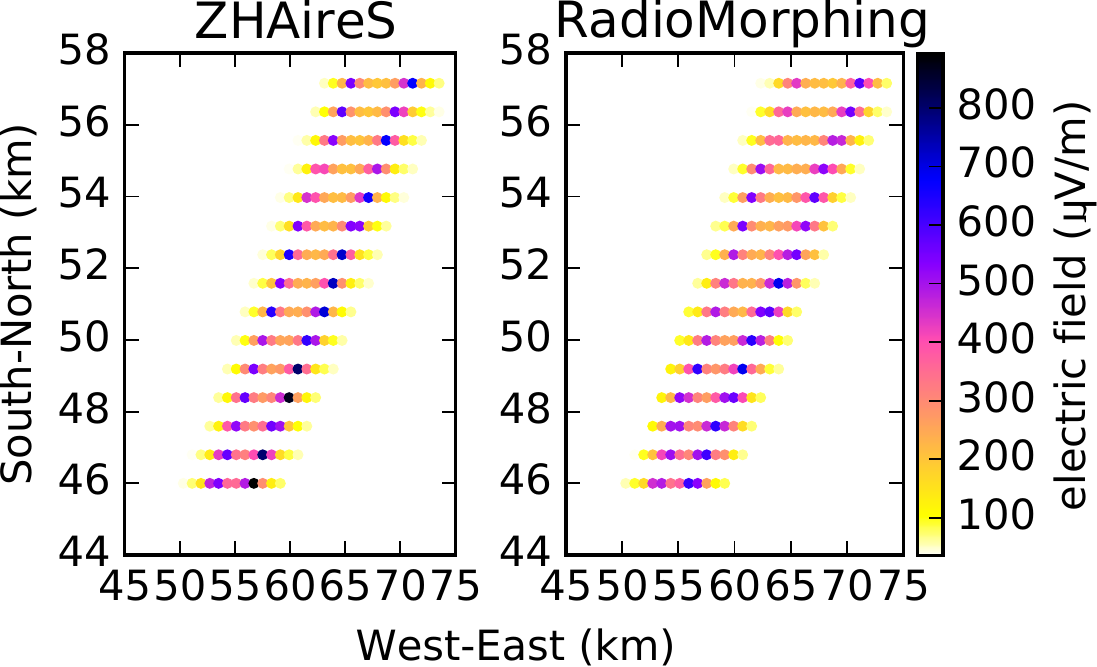}
	\caption{\label{fig:example_2D} Comparison of the footprint detected by an antenna array, tilted by $5^\circ$ towards South: the peak amplitude distributions once simulated using ZHAireS (left) and once calculated by Radio Morphing (right) for the example target shower.}
\end{figure}

Figure~\ref{fig:example_2D} shows the peak-amplitude distribution of received signals (West-East components) emitted by the same example target shower, simulated using ZHAireS (left) and calculated with Radio Morphing (right).
The observer positions are located at a distance of roughly $d=75\,$km from the injection point of the shower on a slope slightly tilted by $5^\circ$ towards South. The antenna positions are arranged in a grid-like structure, as planed for the future GRAND experiment.
The direct comparison shows that for an extended array, the predictions for the signal strength by Radio Morphing and ZHAireS are in good agreement. The feature of the Cherenkov cone is clearly visible in both distributions.

\begin{figure*}[tb] 
	\centering
		\includegraphics[width=0.8\textwidth]{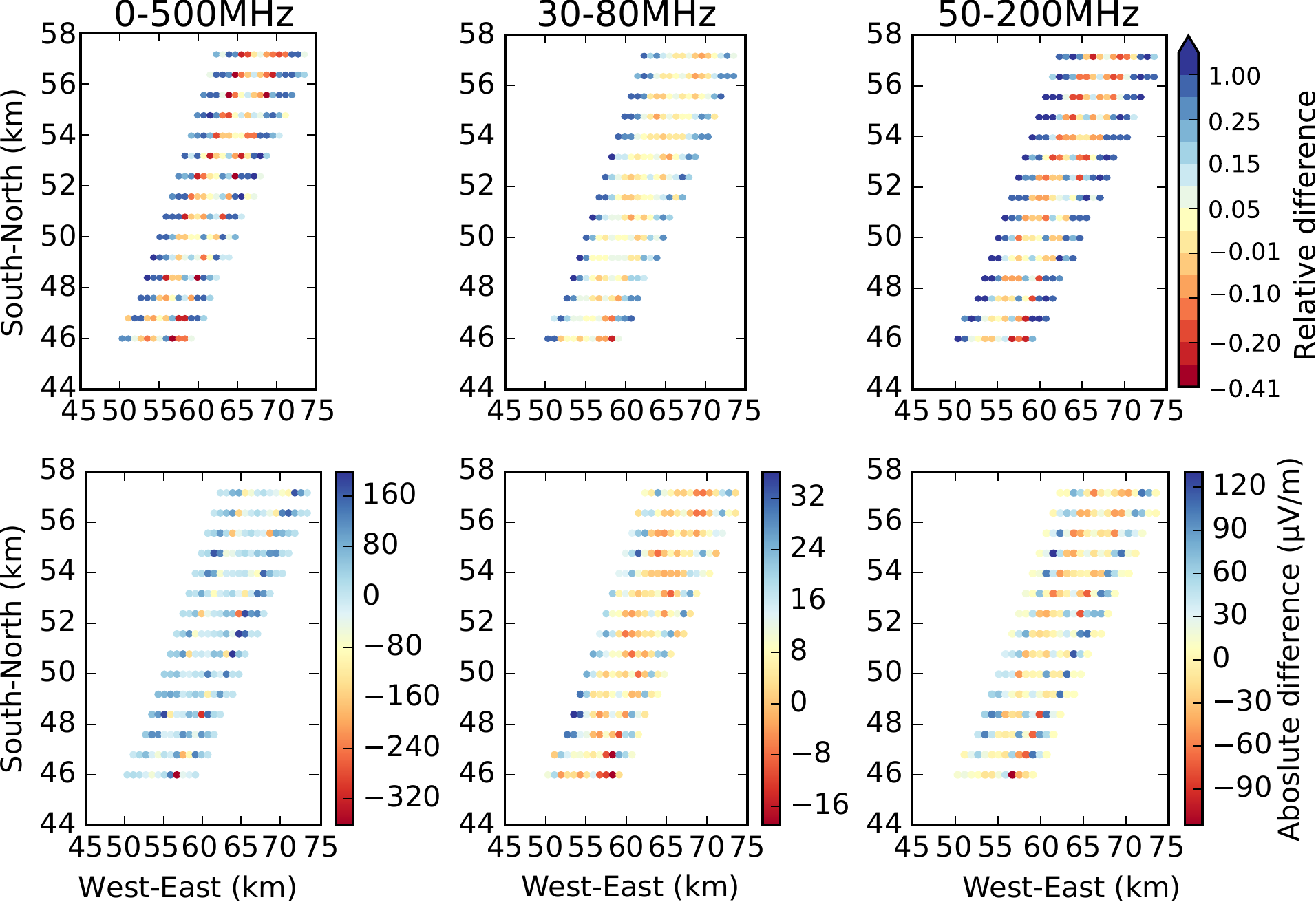}
	\caption{\label{fig:Diff_2D} The relative (top) and the absolute (bottom) differences in the West-East component of the signal distribution, defined as $(E_{\rm{RM}} - E_{\rm{ZHAireS}}) /E_{\rm{ZHAireS}}$ and $E_{\rm{RM}} - E_{\rm{ZHAireS}}$, for the full frequency band of $0-500\,$MHz (left) and for the frequency bands $30-80\,$MHz (center) and $50-200\,$MHz (right).}
\end{figure*}

Figure~\ref{fig:Diff_2D} presents the relative and absolute differences between ZHAireS and Radio Morphed footprints for this same example shower, defined as $(E_{\mathrm{RM}} - E_{\mathrm{ZHAireS}}) /E_{\mathrm{ZHAireS}}$ (top) and  $E_{\mathrm{RM}} - E_{\mathrm{ZHAireS}}$ (bottom), for the full frequency band of $0-500\,$MHz (left) and for the frequency bands $30-80\,$MHz (center) and $50-200\,$MHz (right). One can observe that the highest differences in the peak-amplitude distributions appear at the edges of the Cherenkov ring. This corresponds to the slight mismatch in the predicted positions of the Cherenkov cone discussed earlier. Since the signal strength drops sharply slightly off the cone angle, it leads to a larger difference in the predicted signal strength.

One observes that the signal predicted by Radio Morphing is slightly underestimated for observer positions on the Cherenkov ring, while it is slightly overestimated outside. This is induced mainly by the choice of the reference shower in this specific example, since a low-energy air shower with an energy of $0.1\,$EeV with a flat lateral distribution function acts as the reference for a target shower with an energy of $1.05\,$EeV. The discrepancy in the signal strength decreases if the time traces are filtered, as done for the $30-80\,$MHz (center) and $50-200\,$MHz (right) bands. Here, one first reason is the numerical noise at higher frequencies due to a low thinning level. Besides, coherence conditions are not fulfilled for frequencies above $\sim 100\,$MHz, so that the linear scaling in energy induces larger uncertainties at higher frequencies.

\subsection{Statistical validation of Radio Morphing}\label{sec:statvalid}

\begin{figure*}[tb] 
	\centering
		\includegraphics[width=0.65\textwidth]{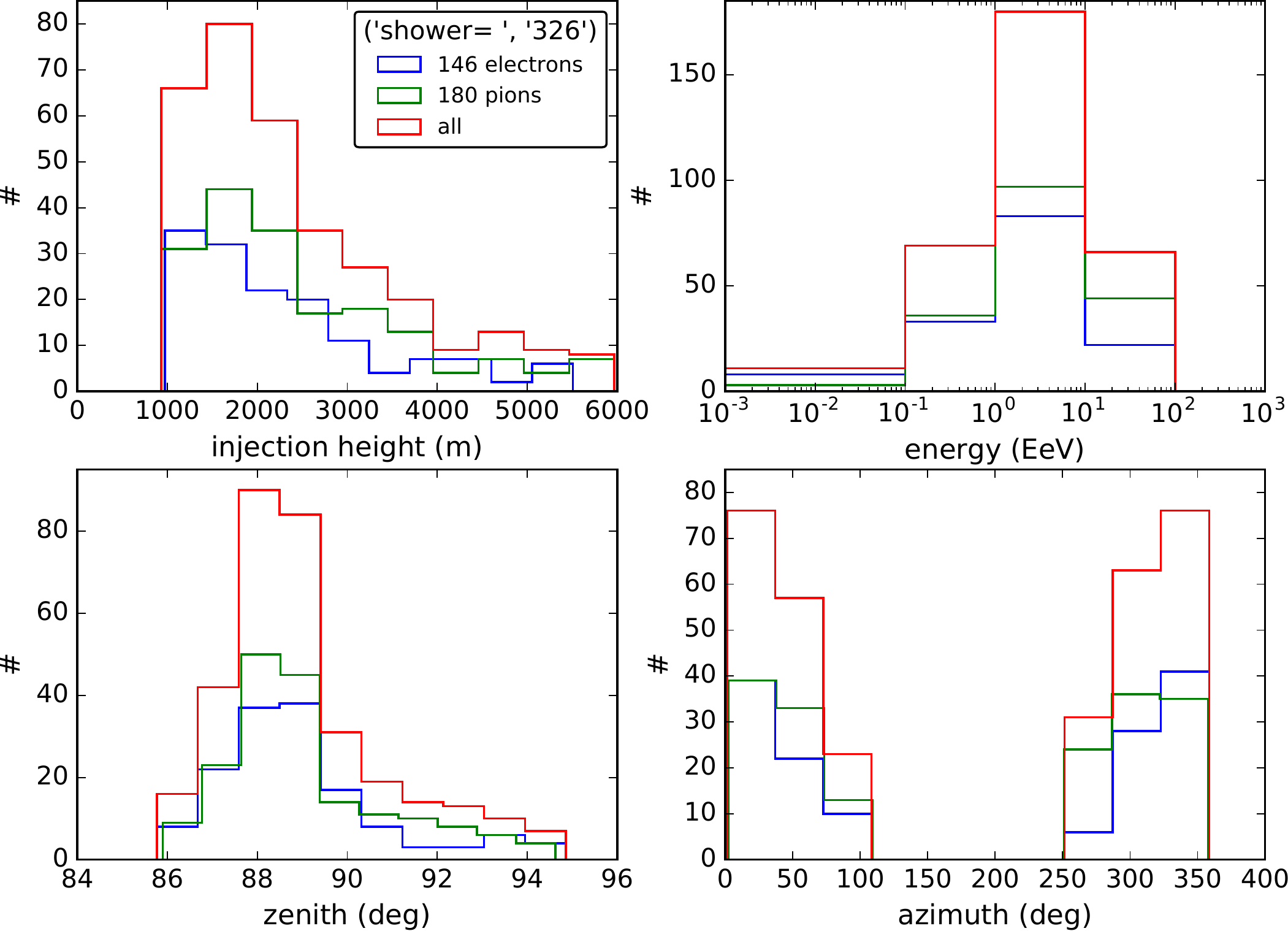}
	\caption{\label{fig:eventset} Characterization of shower events in the set: distribution of injection heights, primary energies, azimuth and zenith angles of the pion (green) and electron (blue) primaries inducing the target air showers. The red line describes the distributions of all events in the set.}
\end{figure*}

\begin{figure*}[tb] 
	\centering
		\includegraphics[width=0.65\textwidth]{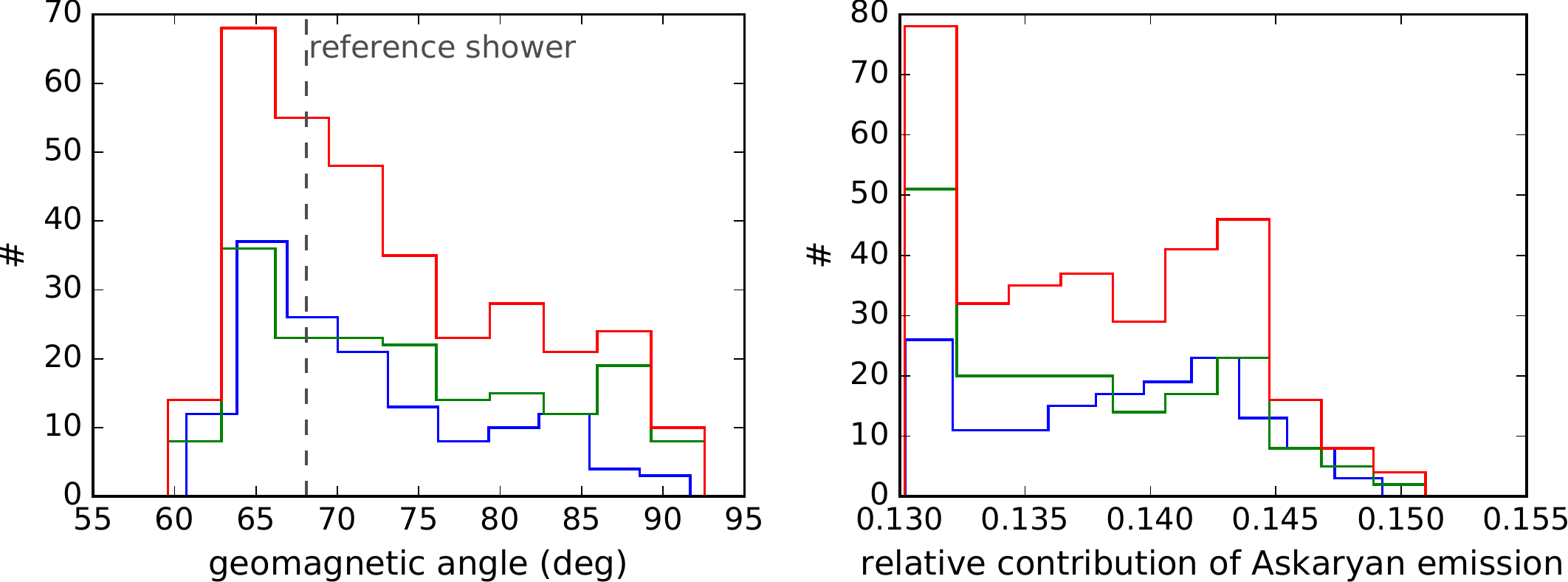}
	\caption{\label{fig:eventset_geo} Left: Determined geomagnetic angles for the events shown in Fig.~\ref{fig:eventset}. The vertical dashed line marks the value for the used reference shower. Right: The derived contribution of the Askaryan emission to the maximal amplitude for an observer position at the Cherenkov cone.}
\end{figure*}

In the following, a comparison is made between results from Radio Morphing and ZHAires. The comparison is based on a set of $\sim300\,$inclined air showers induced by high-energy electrons and pions that are propagating towards the North. The distribution in injection height above sea level, energy, zenith and azimuth are shown in Figure~\ref{fig:eventset}. Figure~\ref{fig:eventset_geo} shows the corresponding values for the geomagnetic shower and the determined contribution of the Askaryan emission to the peak amplitude at an observer position on the Cherenkov cone. For the reference shower, we can decouple the contribution from the Geomagnetic and the Askayran emission for the simulated observer position along the positive $\vec{v}\times\vec{B}$-axis due to the polarisation of the two signal contributions. We found an Askaryan contribution of about $14$\% for a position on the Cherenkov cone. Based on the given formula in \cite{14percent}, we calculate the expected relative contribution to the signal strength at this position by scaling with the geomagnetic angle. The overall assumption that the impact of the Askaryan emission is minor is also valid for the used event set.
We computed the position of the shower maximum $X_{\mathrm{max}}$ of the target showers as described in Section \ref{section:isometries}. Here, we obtained the value for $X_{\mathrm{max}}$ in $\mathrm{g}.\mathrm{cm}^{-2}$ from ~\cite{Risse:2009ir}, while we determine the atmospheric column depth between the injection point and $X_{\mathrm{max}}$ of the electron (pion)-induced shower from the elongation rate value of photon (proton)-induced ones.
The radio signals were computed at observer positions located inside the showers radio footprint, typically $40-50\,$km away from the shower injection point. Only observer positions with off-axis angles smaller than  $1.6^\circ$ were considered.

\begin{figure}[tb] 
	\centering
		\includegraphics[width=0.45\textwidth]{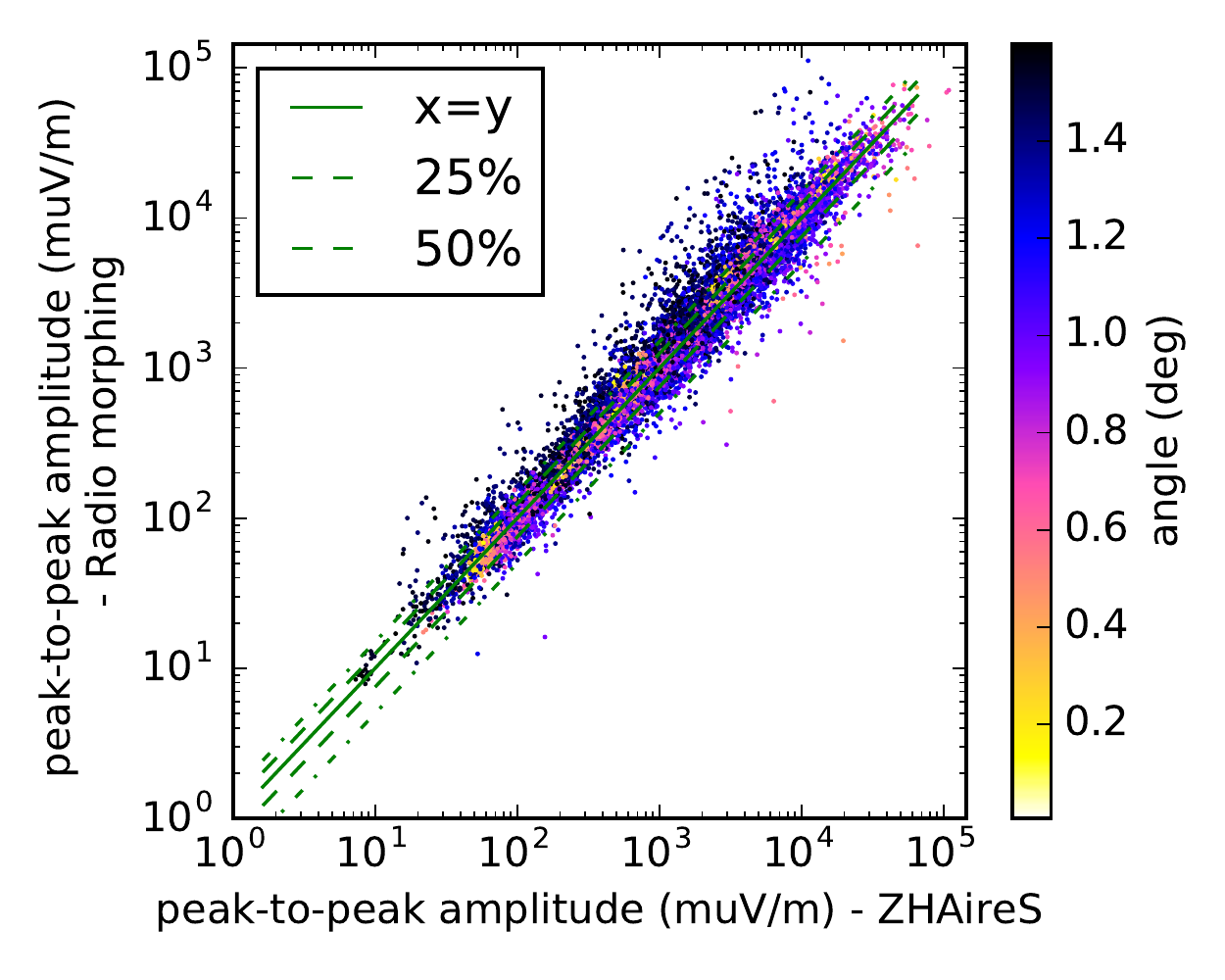} 
	\caption{\label{fig:lead_amp} Comparison of the peak-to-peak amplitude in the East-West component of the signal detected by each single antenna included in the set, calculated with Radio Morphing and simulated with ZHAireS. The color-code represents the location of the observer position with respect to the shower axis, given as the corresponding off-axis angle. The green solid line marks equivalent amplitudes, the dashed line where the results are $25\%$ off, and the dashed-dotted line stands for $50\%$ discrepancy.}
\end{figure}

The same reference shower as for the example above (see Sec. \ref{sec:comp1}) was used to compute the Radio Morphed signals.  
Figure~\ref{fig:lead_amp} displays the calculated peak-to-peak amplitude for each antenna position in the events derived with Radio Morphing versus the peak-to-peak amplitude from ZHAireS simulations. The antenna positions are arranged in a grid-like structure, as also used in Fig.~\ref{fig:example_2D}.

The green solid line marks equivalent amplitudes, the dashed and the dotted-dashed line $25\%$ and $50\%$ offsets, respectively. 
The distribution follows a linear trend, clustering along the diagonal. For most of the positions, the peak-to-peak amplitudes deviate by less than $25\%$. This demonstrates that Radio Morphing can reproduce the amplitudes at the same level of magnitude as ZHAireS.

\begin{figure}[tb] 
	\centering
		\includegraphics[width=0.45\textwidth]{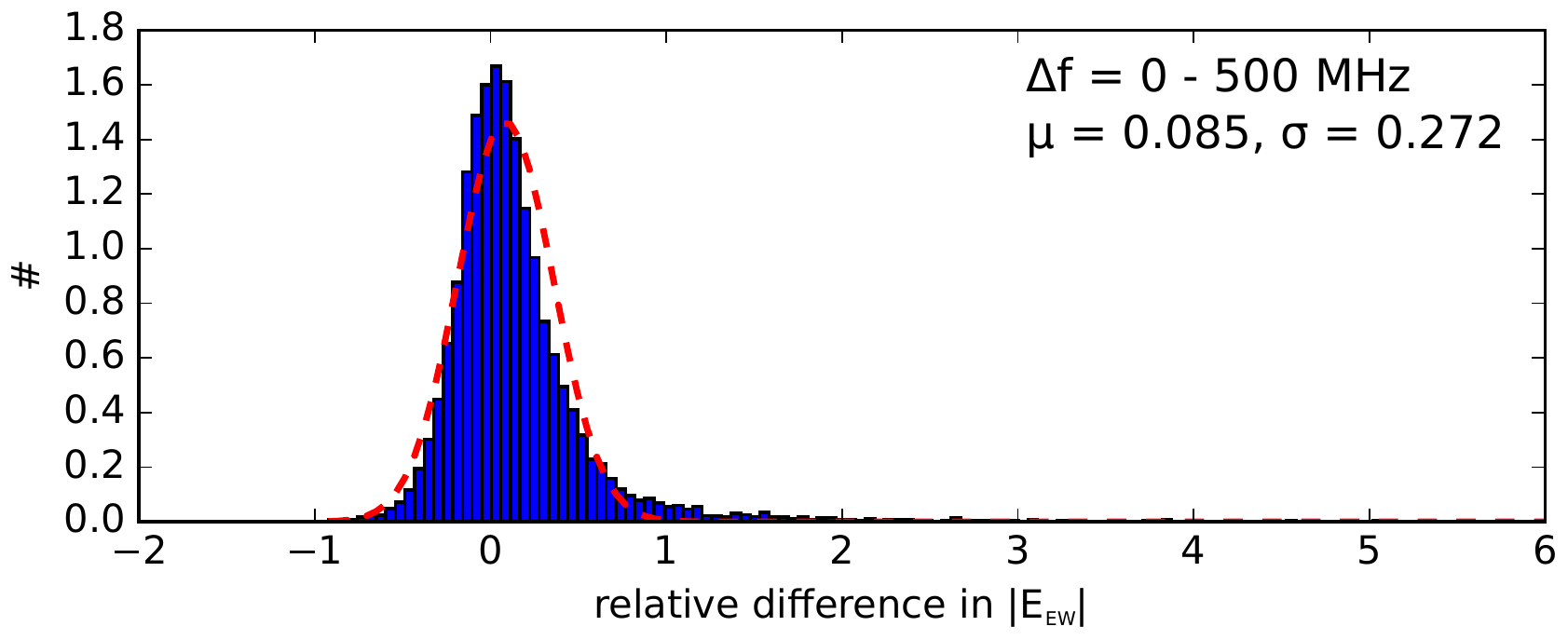}\\
		\includegraphics[width=0.45\textwidth]{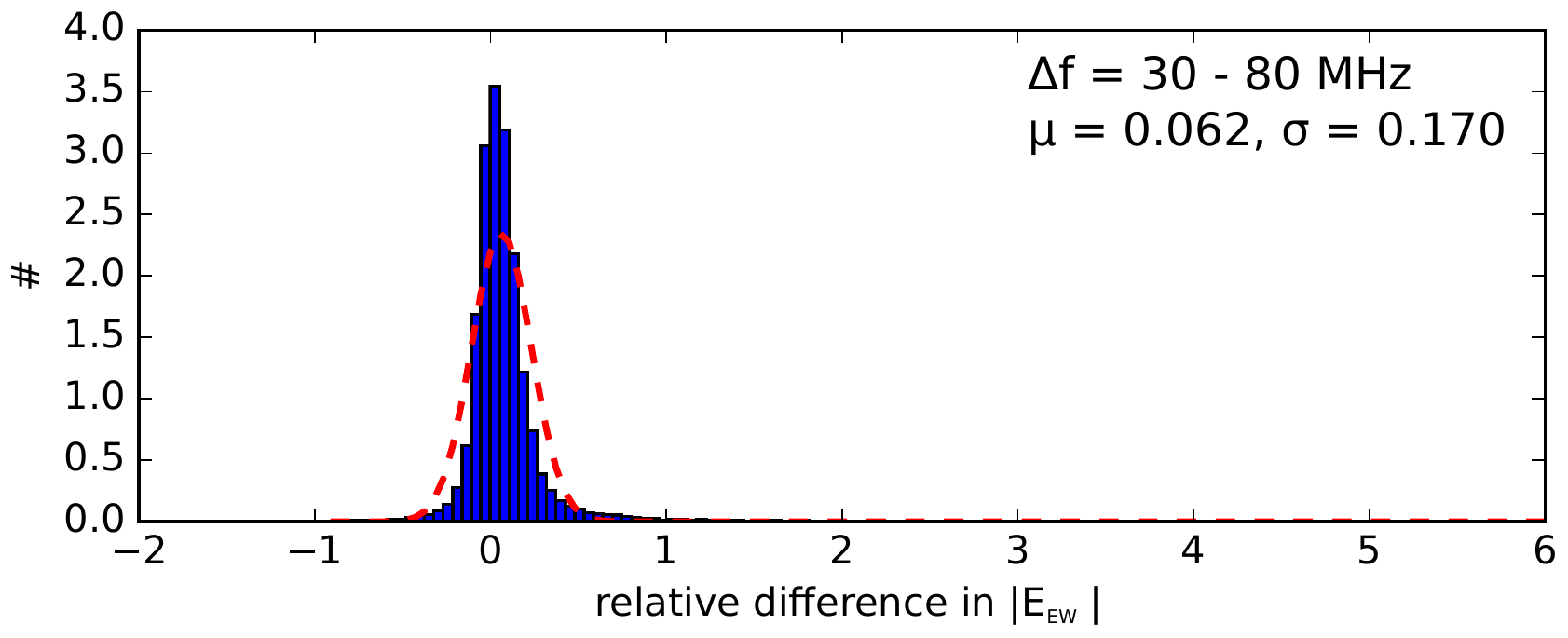}\\
		\includegraphics[width=0.45\textwidth]{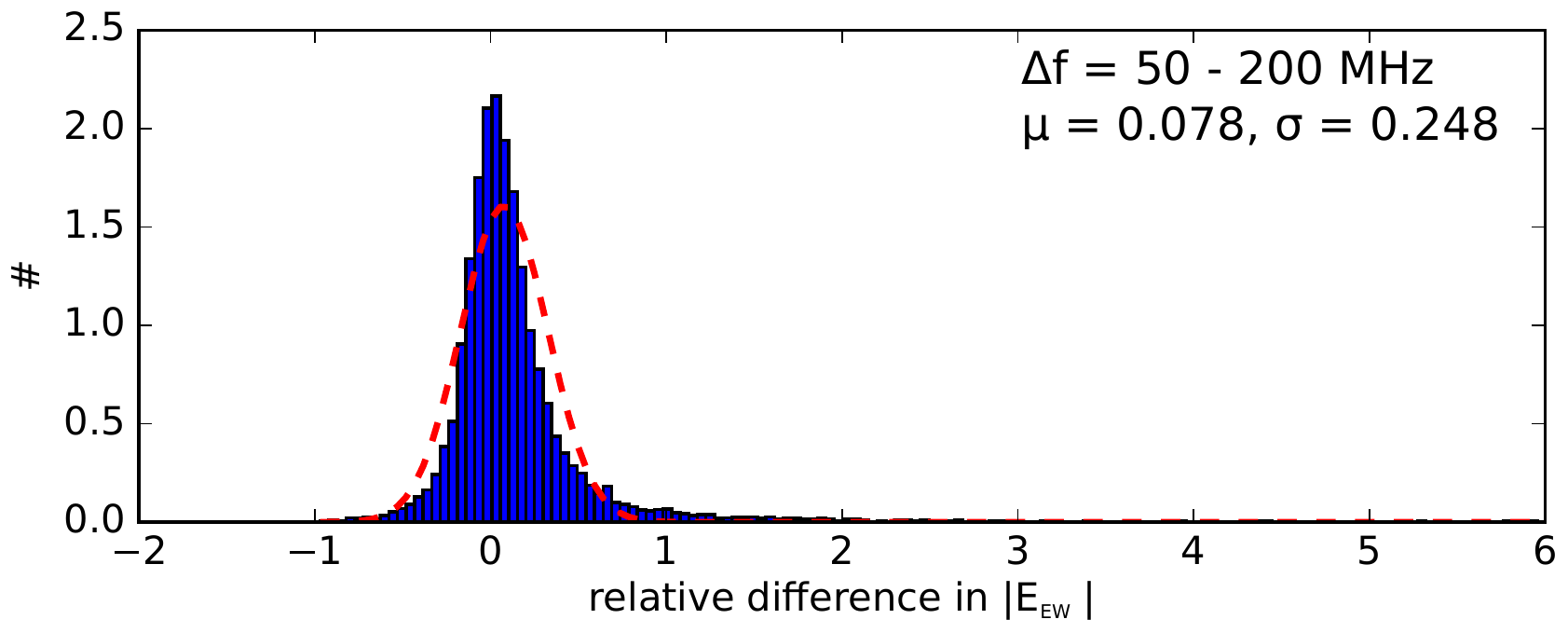}
	\caption{\label{fig:lead_histo} Normalized histograms of the relative difference of the peak-to-peak amplitude at each antenna position in the event set, predicted by Radio Morphing and ZHAireS. For each position, the relative difference is calculated over the full frequency band (top), and over the $30-80\,$MHz (center) and $50-200\,$MHz (bottom) frequency bands. For each distribution a Gaussian function characterized by its mean $\mu$ and its standard deviation $\sigma$ is overlayed. To minimize the impact of extreme outliers, the Gaussian fit was performed only on the data with relative differences $<1$.}
\end{figure}

A histogram of the relative differences between the Radio Morphing and the ZHAireS peak-to-peak amplitudes at each antenna position is presented in Figure~\ref{fig:lead_histo} (top). When excluding the values with relative differences larger than 100\% --- only 1\% of the total --- the mean of the distribution, derived by a Gaussian fit, is $\mu=8.5\%$ with a standard deviation of $\sigma=27.2\%$. The Gaussian function with the mean and the standard deviation is plotted on top for comparison. It appears clearly that the function cannot describe the distribution properly due to a long asymmetric tail towards positive values. These entries arise mainly from the overestimation of the signal by Radio Morphing at antenna positions outside the Cherenkov ring as mention before. The fit does not have a quantitative value, but the plots aim to demonstrate qualitatively that Radio Morphing and ZHAireS agree within a reasonable margin. When the signals are filtered in the $30-80\,$MHz frequency range, then the mean and standard deviations decrease to $\mu=6.2\%$ and $\sigma=17.0\%$ and to $\mu=7.8\%$ and $\sigma=24.8\%$ after a filtering to $50-200\,$MHz, respectively. These behaviors are consistent with our observations in Figure~\ref{fig:Diff_2D}.

\section{Discussion}\label{section:discussion}

The previous section demonstrates the consistency between Radio Morphing and microscopic simulations. Given that the method is built on simple mathematical operations, this agreement is remarkable. 

The major advantage of Radio Morphing compared to microscopic simulations lies in the huge gain in computation time. 
While microscopic simulation programs as ZHAireS require several minutes to calculate the signal for each antenna position (e.g. $\mathcal{O}(\rm mins)$ on one node for a thinning level of $10^{-4}$, ten times more for a thinning level of $10^{-5}$), Radio Morphing requires $\mathcal{O}(\rm s)$ to calculate the electric field trace in the time domain at a desired position, once the reference shower is prepared. Here, the scaling of the amplitude and the positions of the reference shower requires the largest fraction of the computing time and scales linearly with the number of antenna positions included in the reference shower.

\subsection{Radio Morphing systematic uncertainties}

In this section we propose a qualitative discussion on the systematic bias associated with Radio Morphing, and how it can be reduced. We found in the previous section for the given example event set and reference shower a $(8.5\pm27.2)\%$ difference in peak amplitude between signals computed with Radio Morphing and ZHAireS simulation. This discrepancy may obviously impact results when using radio simulation for a specific study, e.g. the trigger rate of a radio array on air showers. This systematic bias may be reduced by two means essentially:

\begin{itemize}
\item increase the numbers of simulated antenna positions in the reference shower. This effectively means a decrease of the distance in-between the simulated antenna positions along the arms in the star-shape pattern and therefore a finer sampling of the shower. That leads to a reduction of the uncertainty in the linear interpolation of the pulse shape based on the plane-wave approximation. In addition, a smaller distance in-between the star-shape planes leads to a more precise sampling of the reference shower and therefore a better sampling of the changes in the field-strength distribution along the direction of propagation.   
Note however that with a rising number of simulated antenna positions, not just the simulation time required for the production of the reference shower increases, but also the scaling operations within the radio-morphing method will last longer. 

\item the uncertainty in the scaling also rises with the difference in the parameters values between the reference and the target showers. It means that more than one reference shower may have to be included in the target shower computation process, depending on the actual parameter range to be covered, application case and desired accuracy. Having more than one reference shower would lead to a reduction of the number of extreme outliers in the electric-field strength distribution (compare to Fig. \ref{fig:lead_histo}). Therefore, the spread in the offset to results obtained by ZHAireS simulations will decrease.

\end{itemize}

Another systematic uncertainty induced by using Radio Morphing is that so far the asymmetry in the signal footprint caused by the superposition of the geomagnetic and Askaryan effect is not yet included in the scaling. This means that the asymmetry in the signal distribution does not depend on the azimuth angle. Just the asymmetry information contained in the reference shower is conserved, and not adjusted for the new target geometry. This effect can as well be weakened by using more than one reference shower in Radio Morphing. The scaling of the signal asymmetry due to the interference of the two main emission mechanisms can be implemented by the disentanglement of the two components in shower coordinates and the identification of the Askaryan contribution by running reference simulations with the magnetic field switched on and off. This will be included in a future version of the Radio Morphing code.

\section{Summary}
{\it Radio Morphing} is a newly developed universal tool to calculate the radio signal of an air-shower, combining the precision of microscopic simulations with the speed of macroscopic approaches.  It consists in simple mathematical operations performed on a reference shower and in simple signal interpolation in the frequency domain. The mathematical operations are based on theoretical and measured parametrizations of the radio signal on the characteristics of the primary particle. The computation speed is independent of thinning level of the air shower, in contrast to microscopic simulations. 

With Radio Morphing, it is possible to achieve an impressive gain in computation time, while reproducing accurately all three electric field components at any antenna position. In particular, features such as the Cherenkov cone and the signal strength at any observer positions are correctly modeled. It is thus an ideal tool to perform fast simulations of non-flat topographies, as required for example, in the calculations of the performances of the GRAND project \cite{Alvarez-Muniz:2018bhp}. 

A more systematic quantification of the relative errors compared to microscopic simulations requires testing on a specific layout and geographical assumptions. This is currently being explored within the framework of the GRAND project. 
Other limitations of the method, such as the time interpolation of the signal, are also under investigation.

\subsection*{Acknowledgments}
This work is supported by the APACHE grant (ANR-16-CE31-0001) of the French Agence Nationale de la Recherche and by the grant \#2015/15735-1, São Paulo Research Foundation (FAPESP).
This work has made use of the Horizon Cluster hosted by Institut d'Astrophysique de Paris. Part of the simulations were performed using the computing resources at the CC-IN2P3 Computing Centre (Lyon/Villeurbanne – France),  partnership between CNRS/IN2P3 and CEA/DSM/Irfu.


\bibliography{bib_rm} 

\end{document}